%% file: nonres_paper.tex
\PassOptionsToPackage{colorlinks,breaklinks}{hyperref}  
\documentclass[epjc3,twocolumn,sort&compress]{svjour3}
\RequirePackage[T1]{fontenc}

\smartqed

\RequirePackage{graphicx}
\RequirePackage{mathptmx}  
\RequirePackage{flushend}
\RequirePackage[numbers,sort&compress]{natbib}

\usepackage[utf8]{inputenc}
\usepackage{subfigure}
\usepackage{multirow}
\usepackage{graphicx}
\usepackage{lineno}
\usepackage{amssymb}
\usepackage{amsmath}
\usepackage[dvipdfm,colorlinks,citecolor=blue,urlcolor=blue,linkcolor=blue]{hyperref}
\usepackage{setspace}
\usepackage{booktabs}
\usepackage{rotating}
\usepackage{lldefs}
\usepackage{rotating}
\usepackage{verbatim}
\usepackage{url}
\usepackage{array} 
\usepackage{textcomp} 
\usepackage{morefloats}
\usepackage{atlasphysics}

\graphicspath{{.}}

\usepackage{preprintcover}
\PreprintCoverPaperTitle{Search for contact interactions and large extra dimensions in the dilepton channel using proton--proton collisions at $\sqrt{s}~=$~8~\TeV{} with the ATLAS detector}  
\PreprintIdNumber{CERN-PH-EP-2014-135}  
\PreprintCoverAbstract{\noindent A search is conducted for non-resonant new phenomena in dielectron and dimuon final states, originating from either contact interactions or large extra spatial dimensions. The LHC 2012 proton--proton collision dataset recorded by the ATLAS detector is used, corresponding to 20~fb$^{-1}$ at $\sqrt{s}$ = 8~\TeV. The dilepton invariant mass spectrum is a discriminating variable in both searches, with the contact interaction search additionally utilizing the dilepton forward-backward asymmetry. No significant deviations from the Standard Model expectation are observed. Lower limits are set on the $\ell\ell q q$ contact interaction scale $\Lambda$ between 15.4~\TeV{} and 26.3~\TeV{}, at the 95\% credibility level. For large extra spatial dimensions, lower limits are set on the string scale $M_{\rm S}$ between 3.2~\TeV{} to 5.0~\TeV{}.}  
\PreprintJournalName{Eur. Phys. J. C}  

\journalname{Eur. Phys. J. C}

\begin{document}

\title{Search for contact interactions and large extra dimensions in the dilepton channel using proton--proton collisions at $\sqrt{s}~=$~8~\TeV{} with the ATLAS detector}

\titlerunning{Search for contact interactions and large extra dimensions in the dilepton channel}

\author{ATLAS Collaboration}

\institute{CERN, 1211 Geneva 23, Switzerland \\ \email{atlas.publications@cern.ch}}

\date{Received: 10 July 2014 / Accepted: 16 October 2014}


\maketitle

\begin{abstract}

\noindent A search is conducted for non-resonant new phenomena in dielectron and dimuon final states, originating from either contact interactions or large extra spatial dimensions. The LHC 2012 proton--proton collision dataset recorded by the ATLAS detector is used, corresponding to 20~fb$^{-1}$ at $\sqrt{s}$ = 8~\TeV. The dilepton invariant mass spectrum is a discriminating variable in both searches, with the contact interaction search additionally utilizing the dilepton forward-backward asymmetry. No significant deviations from the Standard Model expectation are observed.  Lower limits are set on the $\ell\ell q q$ contact interaction scale $\Lambda$ between 15.4~\TeV{} and 26.3~\TeV{}, at the 95\% credibility level. For large extra spatial dimensions, lower limits are set on the string scale $M_{\rm S}$ between 3.2~\TeV{} to 5.0~\TeV{}.

\end{abstract}

%


\section{Introduction}

Many theories beyond the Standard Model (SM) predict new phenomena which give rise to dilepton final states, such as new resonances. These have been searched for using the ATLAS detector at the Large Hadron Collider (LHC) and are reported elsewhere~\cite{Aad:Zprime}. In this paper, a complementary search is performed for new
phenomena that appear as broad deviations from the SM in the dilepton invariant mass distribution or
in the angular distribution of the leptons (where the leptons considered in this analysis are electrons or muons).
The phenomena under investigation are contact interactions (CI) and large extra
dimensions (LED).

\section{Theoretical motivation}

The presence of a new interaction can be detected at an energy
much lower than that required to produce direct evidence of the
existence of a new gauge boson. The charged weak interaction
responsible for nuclear $\beta$ decay provides such an example. 
A non-renormalizable description of this
process was successfully formulated by Fermi
in the form of a four-fermion contact interaction~\cite{Fermi:1934hr}.
A contact interaction can also accommodate deviations from the SM in
proton--proton scattering due to quark and lepton compositeness, where a
characteristic energy scale $\Lambda$ corresponds to the binding
energy between fermion constituents.
A new interaction or compositeness in the process 
$q\overline{q} \to \ell^+\ell^-$ can be described by the following four-fermion
contact interaction Lagrangian~\cite{Eichten:1983hw, Eichten:1984eu}

\vspace{-0.3cm}

\begin{equation}\label{lagrangian}
\begin{array}{r@{\,}c@{}c@{\,}l@{\,}l}
\mathcal L = \frac{g^2}{\Lambda^2}\;[ && \eta_{\rm LL}&\, (\overline q_{\rm L}\gamma_{\mu} q_{\rm L})\,(\overline\ell_{\rm L}\gamma^{\mu}\ell_{\rm L}) \nonumber \\
& +&\eta_{\rm RR}& (\overline q_{\rm R}\gamma_{\mu} q_{\rm R}) \,(\overline\ell_{\rm R}\gamma^{\mu}\ell_{\rm R}) \\
&+&\eta_{\rm LR}& (\overline q_{\rm L}\gamma_{\mu} q_{\rm L}) \,(\overline\ell_{\rm R}\gamma^{\mu}\ell_{\rm R}) \\
&+&\eta_{\rm RL}& (\overline q_{\rm R}\gamma_{\mu} q_{\rm R}) \,(\overline\ell_{\rm L}\gamma^{\mu}\ell_{\rm L})& ] \: ,\nonumber
\end{array}
\end{equation}

\noindent where $g$ is a coupling constant chosen by convention to satisfy $g^2/4\pi = 1$,
$\Lambda$ is the contact interaction scale,
and $q_{\rm L,R}$ and $\ell_{\rm L,R}$ are left-handed and right-handed quark and lepton fields, respectively.
The parameters $\eta_{ij}$, where $i$ and $j$ are L or R (left or right), 
define the chiral structure of the new interaction.
Different chiral structures are investigated here, with the left-right model
obtained by setting $\eta_{\rm LR} = \eta_{\rm RL} = \pm 1$ and $\eta_{\rm LL} = \eta_{\rm RR} = 0$.
Likewise, the left-left and right-right models are obtained by setting the corresponding parameters to $\pm 1$, and the others to zero. The sign of $\eta_{ij}$ determines whether the interference is constructive ($\eta_{ij} = -1$) or destructive ($\eta_{ij} = +1$). The cross-section for the process $q \overline{q} \to \ell^+\ell^-$ in the presence of these contact interaction models can be written as:

\vspace{-0.3cm}

\begin{eqnarray}
\sigma_{\rm tot} = \sigma_{\rm DY} - \eta_{ij} \frac{F_{\rm I}}{\Lambda^2} + \frac{F_{\rm C}}{\Lambda^4}~,
\label{cross_section_CI}
\end{eqnarray}

\noindent where the first term accounts for the $q \overline{q} \to Z/\gamma^* \to \ell^+\ell^-$ 
Drell--Yan (DY) process, the second term corresponds to the interference between the DY
and CI processes, and the third term describes the pure CI process. These two latter terms include $F_{\rm I}$ and $F_{\rm C}$, respectively, which are functions of the cross-section, and do not depend on $\Lambda$.
The relative impact of the interference and pure CI terms
depends on both the dilepton mass and $\Lambda$. For example, the
magnitude of the interference term for dilepton masses above 600~\GeV{} is about twice as large as that of the pure CI term at $\Lambda = 14$~\TeV{}; the interference becomes increasingly dominant for higher values of $\Lambda$.

There are other models which predict deviations from the SM in the dilepton mass spectrum, and seek to address the vast hierarchy between the electroweak (EW) and Planck scales, such as the solution proposed by Arkani-Hamed, Dimopoulos and Dvali (ADD)~\cite{ArkaniHamed:1998rs}.
In this model, gravity is allowed to propagate in large flat extra spatial dimensions,
thereby diluting its apparent effect in 3+1 spacetime dimensions.
The flat $n$ extra dimensions are of common size $R$  
($\sim$1~\textmu m--1~mm for $n$ = 2) 
and are compactified on an 
$n$-dimensional torus. The fundamental Planck scale in (4+$n$)-dimensions,
$M_{\rm D}$, is related to the Planck scale, $M_{\rm Pl}$, by Gauss's law
$M^{2}_{\rm Pl}$ $\sim$ $M_{\rm D}^{n+2} R^n$.
It is thus possible for $M_{\rm D}$ to be in the \TeV{} range for sufficiently large volumes ($\propto R^n$).
In this model, the SM particles and their interactions
are confined to a four-dimensional submanifold, whereas gravitons may also propagate into
extra dimensions of size $R$. This gives rise to a tower of Kaluza--Klein (KK)
graviton modes with a mass spacing inversely proportional to $R$.
Values for $M_{\rm D}$ at the \TeV{} scale imply very small mass
differences between KK modes and thus an essentially continuous mass spectrum. 

The production of dileptons via virtual KK graviton exchange involves a sum
over many KK modes that needs to be cut off at some value. In this
paper, the ultraviolet cutoff is chosen to be the string scale, $M_{\rm S}$~\cite{Witten:1996mz}, which sets the context in which this search and its results should be interpreted, and is chosen for consistency with previous searches.
This scale is related to $M_{\rm D}$ via the Gamma function, $\Gamma$, by~\cite{Gleisberg:2003ue}

\vspace{-0.3cm}

\begin{equation}
M_{\rm S} = 2 \sqrt{\pi} \left[ \Gamma\left( \frac{n}{2} \right) \right]^{1/(n+2)} M_{\rm D}~.\nonumber
\label{EqMSMD}
\end{equation}

The cross-section for $q \overline{q}/gg \to \ell^+\ell^-$ in the presence of large extra
dimensions can be expressed as

\vspace{-0.3cm}

\begin{eqnarray}\label{cross_section_ADD}
\sigma_{\rm tot} =  \sigma_{\rm DY} + \mathcal{F}\frac{F_{\rm int}}{M_{\rm S}^4} + \mathcal{F}^{2}\frac{F_{\rm G}}{M_{\rm S}^8},
\end{eqnarray}

\noindent where $\sigma_{\rm DY}$ is the DY cross-section, and
$F_{\rm int}$ and $F_{\rm G}$ are functions of the cross-sections (they do not depend on $M_{\rm S}$) involving the interference and pure KK graviton effects, respectively.
The strength of the interaction is characterized by 
$\mathcal{F}$/$M^{4}_{\rm S}$, where the dimensionless
parameter $\mathcal{F}$ varies in the different calculations provided by 
Giudice--Rattazzi--Wells (GRW)~\cite{Giudice:1998ck}, Hewett~\cite{Hewett:1998sn} 
and Han--Lykken--Zhang (HLZ)~\cite{Han:1998sg}. The different values are:

\vspace{-0.3cm}

\begin{eqnarray}\label{ADDformalism}
\mathcal F & = & 1,  {\rm ~~(GRW)}  \nonumber \\
\mathcal F & = & \frac{2 \lambda}{\pi} = \frac{\pm 2}{\pi}, {\rm ~~(Hewett)} \nonumber \\
\mathcal F & = & \log \left( \frac{M_{\rm S}^2}{s} \right) {\rm ~for~} n = 2, {\rm ~~(HLZ)} \nonumber \\
\mathcal F & = & \frac{2}{n-2} {\rm ~for~} n > 2. {\rm ~~(HLZ)}. \nonumber
\end{eqnarray}

In the Hewett formalism, $\lambda = \pm 1$ is introduced to allow for
constructive or destructive interference with the DY process.
Unlike the situation with contact interactions described above, interference
effects between the DY and virtual KK graviton processes are small due
to dilepton production by virtual KK gravitons being predominantly gluon-induced rather than quark-induced.

Previous searches for CI have been carried out in neutrino--nucleus and
electron--electron scattering~\cite{Anthony:2005pm}, as well as
electron--positron~\cite{Abdallah:2008ab, Schael:2006wu},
electron--proton~\cite{Aaron:2011mv}, and
proton--antiproton colliders~\cite{Abulencia:2006iv,Abazov:2009ac}.
Searches for CI have also been performed by the ATLAS and CMS
Collaborations~\cite{Aad:2012bsa, Chatrchyan:2012hda}.
The strongest exclusion limits for $\ell\ell q q$ CI in which all quark flavours contribute come from the previous ATLAS non-resonant dilepton analysis conducted using 5~\ifb{} of proton--proton ($pp$) collision data at $\sqrt{s}$ = 7~\TeV{}~\cite{Aad:2012bsa}. That combined analysis of the dielectron and dimuon channels set lower limits at 95\% credibility level (C.L.) on the left-left model of $\Lambda$ $>$ 13.9~\TeV{} and $\Lambda$ $>$ 10.2~\TeV{}, for constructive and destructive interference, respectively, given a uniform positive 1/$\Lambda^2$ prior.

Previous searches for evidence of ADD-model extra dimensions via virtual KK graviton exchange have been performed at electron--positron~\cite{Abbiendi:2002je}, electron--proton~\cite{Chekanov:2003pw}, and
proton--antiproton colliders~\cite{Abazov:2009ac}.
Searches have also been performed at the LHC by the ATLAS and CMS Collaborations~\cite{Chatrchyan:2012kc, Aad:2012bsa}. 
The most stringent results come from the ATLAS search in the dilepton channel and subsequent combination with the diphoton channel result using 5~\ifb{} of $pp$ collision data at $\sqrt{s}$ = 7~\TeV{}~\cite{Aad:2012bsa}. That analysis set lower limits on $M_{\rm S}$ at 95\% C.L. in the GRW formalism of 3.5~\TeV{} and 3.4~\TeV{} for 1/$M_{\rm S}^4$ and 1/$M_{\rm S}^{8}$ priors, respectively.

\section{The ATLAS detector}

The ATLAS detector~\cite{Aad:2008zzm} consists of an Inner Detector (ID) surrounded by a solenoid magnet for tracking charged particles, and a calorimeter
for capturing particles that interact electromagnetically or hadronically, to measure their energy.
A Muon Spectrometer (MS) and toroidal magnet system provide tracking for
muons, which typically escape the calorimeter.

The ID is immersed in a 2.0~T axial magnetic field and provides charged-particle tracking up to $|\eta|$ of 2.5.\protect\footnote{ATLAS uses a right-handed coordinate system with its origin at the nominal
interaction point (IP) in the centre of the detector and the $z$-axis along
the beam pipe. The $x$-axis points from the IP to the centre of the LHC ring,
and the $y$-axis points upward. Cylindrical coordinates $(r,\phi)$
are used in the transverse plane, $\phi$ being the azimuthal angle around
the beam pipe. The pseudorapidity is defined in terms of the polar
angle $\theta$ as $\eta$~=~$-$~$\ln\tan(\theta /2)$.}
It is composed of a pixel detector, a silicon-strip tracker, and a transition
radiation tracker. 

The calorimeter system surrounds the solenoid and extends up to $|\eta|$~=~4.9.
One of its main components is a lead and liquid-argon electromagnetic sampling calorimeter, covering $|\eta|$~$<$~3.2 with a fine segmentation varying by layer.
This provides precise energy and position measurements for electrons
and photons.
Another electromagnetic calorimeter, in the forward direction up to $|\eta|$~=~4.9, uses liquid-argon active elements and copper as an absorber.
Further from the interaction point lies an iron and scintillator tile calorimeter
up to $|\eta|$~=~1.7 and a copper and liquid-argon calorimeter up to $|\eta|$~=~3.2
for hadronic energy measurements.
A hadronic calorimeter in the forward region, up to $|\eta|$~=~4.9, uses liquid-argon active elements combined with tungsten as an absorber.

The outermost detector is the MS, which consists of layers of
precision tracking chambers and trigger chambers to enable reconstruction of muons
with $|\eta|$~$<$~2.7. Precision tracking is provided by monitored drift tube
chambers, complemented by a layer of cathode strip chambers in the innermost
layer in the forward region.
Triggering is handled by resistive plate chambers in the barrel ($|\eta|$~$<$~1.05) 
and thin-gap chambers in the endcap (1.05~$<$~$|\eta|$~$<$~2.4).
One barrel and two endcap toroidal magnet systems provide the bending force
to measure muon momentum.

The triggering of events to be recorded by the ATLAS detector is handled by a three-level system~\cite{Aad:2008zzm} which consists of a level-1 hardware trigger, and the high-level trigger (HLT). The HLT is made up of the level-2 trigger, which uses regions of interest, and the event filter, which is based on standard ATLAS event reconstruction and analysis algorithms.


\section{Data and Monte Carlo samples}\label{DataMCSamples}

This search uses the LHC 2012 dataset from $pp$ collisions at $\sqrt{s}$ = 8~\TeV{}, corresponding to an integrated luminosity of approximately 20~\ifb{}. The peak luminosity during this period was $7.7 \times 10^{33}~$cm$^{-2}$s$^{-1}$, with an average number of $pp$ interactions per bunch crossing (pile-up) of $\langle$\textmu$\rangle$ = 20.7.

The main background comes from the irreducible DY process.  The photon-induced (PI) process is also an irreducible contribution which produces two leptons, arising from a $\gamma\gamma$ initial state via $\hat{t}$-- and $\hat{u}$-- channel processes.  The PI process is not a major background, although it is important in the description of the lepton angular distribution. 
The reducible, but non-negligible, backgrounds are $t\bar{t}$ and single top-quark production, multi-jet, $W$+jets, and diboson ($WW$, $WZ$, and $ZZ$) processes. Monte Carlo (MC) simulation is used to estimate all of these backgrounds, with the exception of the multi-jet and $W$+jets backgrounds, which are estimated with a data-driven fake-factor method, as described in Sect.~\ref{SectionSelection}. The multi-jet and $W$+jets backgrounds are found to be negligible in the dimuon channel~\cite{Aad:Zprime}.

All MC samples were passed through a simulation of the ATLAS detector using {\sc Geant}4~\cite{Agostinelli:2002hh, Allison:2006ve, G4simulation}. The DY background is generated with \powheg~\cite{Alioli:2010xd} for the next-to-leading-order (NLO) matrix elements using the CT10~\cite{Lai:2010vv} parton distribution functions (PDF) and \pythia~8.165~\cite{Sjostrand:2007gs} for parton showering and hadronization. To correct the DY cross-section from NLO to next-to-next-to-leading-order (NNLO), a dilepton mass-dependent QCD+EW $K$-factor is calculated with \fewz~3.1~\cite{Li:2012wna} using the MSTW2008NNLO~\cite{Martin:2009iq, Martin:2009bu} PDF (with CT10 as the base NLO PDF) to take into account higher-order QCD and EW corrections. The photon-induced background is generated with {\sc Pythia}~8.165 at LO using the MRST2004QED~\cite{Martin:2004dh} PDF. The top-quark production processes are simulated using \mcatnlo~4.06~\cite{Frixione:2002ik} with the CT10 PDF to generate the matrix elements, \jimmy~4.31~\cite{Butterworth:1996zw} to describe multiple parton interactions, and \herwig~6.520~\cite{Corcella:2000bw}  to describe the remaining underlying event and parton showers. Higher-order corrections are calculated with Top$++$~2.0~\cite{Czakon:2011xx} to derive a $K$-factor which scales this background description from NLO to NNLO in QCD, including resummation of next-to-next-to-leading-logarithmic (NNLL) soft gluon terms. The diboson processes are generated with \herwig~6.520 at leading-order (LO) using the CTEQ6L1 PDF~\cite{Pumplin:2002vw}, and these cross-sections are extrapolated to NLO using dilepton mass-independent $K$-factors. 

The CI signal processes are generated using \pythia~8.165 at LO with the MSTW2008LO PDF. The CI cross-section is scaled from LO to NNLO, again using FEWZ with the MSTW2008NNLO PDF to calculate a dilepton mass-dependent QCD+EW $K$-factor. The ADD LED signal process is simulated with the multi-leg LO generator \sherpa~1.3.1~\cite{Gleisberg:2003ue} using the CTEQ6L1 PDF. No higher-order correction is applied to the ADD LED cross-section.

To ensure adequate modelling of the data by the MC simulation, 
data-derived corrections are applied to the simulation.  
These include electron energy scale corrections~\cite{Aad:2014fxa}, muon momentum corrections~\cite{ATLAS-CONF-2013-088}, and pile-up corrections. They also include trigger, lepton identification, and reconstruction scale factors~\cite{Aad:2014fxa, ATLAS-CONF-2013-088}, which are all found to be very close to unity. A summary of the generator, parton shower, and PDF information used for all signal and background MC samples used in this search can be found in Table~\ref{MCTable}.

\begin{table*}[htbp]
\newcommand\T{\rule{0pt}{2.6ex}}
\newcommand\B{\rule[-1.2ex]{0pt}{0pt}}
\caption{Summary of MC sample information for signal and background processes used in this search. The columns from left to right give the process of interest, generator, matrix-element order, parton shower program, and PDF utilized.}
\label{MCTable}
\footnotesize
\centering
\vspace{0.2cm}
\begin{tabular}{c|c|c|c|c}
\hline
\hline
Process \T & Generator & Order & Parton Shower / Hadronization & PDF \\
\hline
$q\bar{q} \rightarrow Z/\gamma^* \rightarrow \ell^+\ell^-$ \T & \powheg~\cite{Alioli:2010xd} & NLO & \pythia~8.165~\cite{Sjostrand:2007gs} & CT10~\cite{Lai:2010vv} \\
$\gamma\gamma/\gamma q/\gamma \bar{q} \rightarrow \ell^+\ell^-$ & \pythia~8.165~\cite{Sjostrand:2007gs} & LO & \pythia~8.165~\cite{Sjostrand:2007gs} & MRST2004QED~\cite{Martin:2004dh} \\
$t\bar{t} \rightarrow \ell X$, $Wt \rightarrow X$ & \mcatnlo~4.06~\cite{Frixione:2002ik} & NLO & \jimmy~4.31~\cite{Butterworth:1996zw} $+$ \herwig~6.520~\cite{Corcella:2000bw} & CT10~\cite{Lai:2010vv} \\
$WW,WZ,ZZ \rightarrow \ell X / \ell \nu / \ell \ell$ \B & \herwig~6.520~\cite{Corcella:2000bw} & LO & \herwig~6.520~\cite{Corcella:2000bw} & CTEQ6L1~\cite{Pumplin:2002vw} \\
\hline
CI: $q\bar{q} \rightarrow \ell^+\ell^-$ \T & \pythia~8.165~\cite{Sjostrand:2007gs} & LO & \pythia~8.165~\cite{Sjostrand:2007gs} & MSTW2008LO~\cite{Martin:2009iq, Martin:2009bu} \\
ADD: $q\bar{q}/gg \rightarrow G^* \rightarrow \ell^+\ell^-$ \B & \sherpa~1.3.1~\cite{Gleisberg:2003ue} & LO (multi-leg) & \sherpa~1.3.1~\cite{Gleisberg:2003ue}  & CTEQ6L1~\cite{Pumplin:2002vw} \\
\hline
\hline
\end{tabular}
\end{table*}


\section{Event selection and background estimation}\label{SectionSelection}
  
Events in the $ee$ channel are required to have passed a two-object trigger with transverse momentum ($p_{\rm T}$) thresholds of 35~\GeV{} and 25~\GeV{}. Events in the $\mu\mu$ channel are required to have passed at least one of two single-object triggers with $p_{\rm T}$ thresholds of 36~\GeV{} and 24~\GeV{}. The higher threshold trigger is used to recover small efficiency losses due to the online muon isolation requirement imposed by the lower threshold trigger. The overall efficiency for dilepton events to fire either of these triggers is found to be $>$ 99\%. In both channels, events are required to have at least one primary vertex with more than two tracks.

In the dielectron channel, events are retained if at least two electrons fulfil the following criteria: 
the electrons satisfy $|\eta|$ $<$ 2.47 (excluding the transition region between the barrel and endcap, 1.37 $<$ $|\eta|$ $<$ 1.52), the leading and sub-leading electrons have $p_{\rm T}$ $>$ 40~\GeV{} and 30~\GeV{}, respectively, and the electrons satisfy a set of electron identification criteria which are designed to reject jets misidentified as electrons~\cite{Aad:2014fxa}. For the leading and sub-leading electrons, the calorimeter isolation must be less than $(0.007 \times E_{\rm T}) + 5.0$~\GeV{}, and $(0.022 \times E_{\rm T}) + 6.0$~\GeV, respectively (where $E_{\rm T}$ is the transverse energy in units of \GeV{}). The electron calorimeter isolation is calculated as the $\sum{E_{\rm T}}$ in a cone of $\Delta R = \sqrt{(\Delta\eta)^2 + (\Delta\phi)^2}$ = 0.2, excluding the electron $E_{\rm T}$. This measure of isolation is corrected for $E_{\rm T}$-dependent leakage, and pile-up effects which are parameterised as a function of the number of primary vertices in the event.
If more than one electron pair exists in the event, the one with the largest scalar sum of $E_{\rm T}$ is chosen. The two electrons in the selected pair are then required 
to have opposite charge and have dilepton mass greater than 80~\GeV.

In the dimuon channel, events are retained if at least two muons fulfil the following criteria: the muons have $p_{\rm T} >$ 25~\GeV, pass track quality requirements, and meet longitudinal ($|z_0|$ $<$ 1~mm) and transverse ($|d_0|$ $<$ 0.2~mm) track impact parameter requirements with respect to the primary vertex. Muons are also required to be isolated: the $\sum{p_{\rm T}}$ of all additional tracks within $\Delta R = 0.3$ of the muon must be less than 5\% of the muon $p_{\rm T}$.
Muons are reconstructed by combining tracks from both the ID and MS systems. 
The MS hit requirements are particularly stringent to improve the momentum
resolution, minimize tails in the dimuon mass distribution, and improve modelling 
by the simulation.
Muon tracks are required to include at least three hits in each of three precision MS chambers and have at least one hit in the non-bending plane ($\phi$) of two separate chambers to determine the $\phi$ coordinate and thus a good estimate of the non-uniform toroidal magnetic field.
If those tracks include hits in precision chambers that have either no alignment or poor alignment, the tracks are rejected.
Finally, the independent ID and MS track $q/p_{\rm T}^{\rm {track}}$ must agree within five standard deviations of the standalone measurement uncertainties added in quadrature.  The muon acceptance is highest in the pseudorapidity region up to approximately 2.5.
If more than two muons satisfy these criteria, the pair of oppositely charged muons with the highest scalar sum of $p_{\rm T}$ is selected. 
The final requirement is that the dimuon mass must be greater than 80~\GeV.

The event selection detailed above is applied to the data and all MC background samples. The acceptance times efficiency for DY events with dilepton mass of 1~\TeV{} (2~\TeV{}) is found to be 67\% (67\%) in the dielectron channel and 47\% (45\%) in the dimuon channel. The selection efficiency is lower for the dimuon channel
mainly because of the strict MS hit requirements.

The dominant DY background, as well as the PI background, $t\bar{t}$ and single top-quark production processes, and diboson processes, are all modelled with MC as described in Sect.~\ref{DataMCSamples}. The combined multi-jet \& $W$+jets background which only affects the electron channel is estimated using a data-driven method designed to describe events which contain a maximum of one real lepton, and one or more jets or photons which are misidentified as a lepton. The details of this method are provided in Ref.~\cite{Aad:Zprime}. For the top and combined multi-jet \& $W$+jets backgrounds, fits are used to describe the shape of the background dilepton mass distribution with a phenomenologically motivated three-parameter ($p_1$, $p_2$, $p_3$) function ($y(x) = p_1\:x^{p_2+p_3 \log x}$, where $x =$ m$_{\ell\ell}$) at high masses, where the statistical uncertainty becomes large. The fit to the top-quarks background is performed in a similar manner to Ref.~\cite{Aad:Zprime} using the mass range 200--700~\GeV{} to match the fit to the MC distribution, and the resultant fit as the extrapolated top background estimate above 500~\GeV{}. The choice of extrapolation point is found to have a negligible effect on the fit; however, the range of the fit and the uncertainty on the fit parameters are included in the systematic uncertainty. For the fit describing the combined multi-jet \& $W$+jets background at high mass, the lower edge of the fit range is varied from 425~\GeV{} to 600~\GeV{} and the upper edge from 700~\GeV{} to 1200~\GeV{}, with the extrapolation point fixed to 500~\GeV{}.  The uncertainty on this fit is negligible compared to the systematic uncertainty assigned to the data-driven method, as described in Sect.~\ref{SectionSystematics}.

In this analysis, the normalization, control, and search regions are defined based on the dilepton mass. In the normalization region with mass between 80~\GeV{} and 120~\GeV{},  the total background estimate is scaled to data. This protects the analysis against mass-independent systematic uncertainties. The control region, defined by the mass range from 120~\GeV{} to 400~\GeV{}, is used to check the quality of the background modelling since the signal contribution is negligible in this region. After the normalization procedure, good agreement is found in the control region, as displayed in Fig.~\ref{Mass}. The small deviation observed in the first bin of the dielectron mass distribution (Fig.~\ref{Mass}) corresponds to an effect that is less than 0.2\% of the total number of events in the normalization region.  Thus it has a negligible effect. The search is then conducted in the mass region 400--4500~\GeV{}.

\section{Event yields}\label{EventYieldSection}

In the CI search, six broad dilepton mass bins are used in the search region from 400~\GeV{} to 4500~\GeV{}. For the ADD search region, a single dilepton mass bin is employed in the range 1900--4500~\GeV{}, where the lower mass boundary is optimized based on the strongest expected exclusion limit.

The dielectron (dimuon) channel event yields are presented in Table~\ref{Electron_Yields_CI} (Table~\ref{Muon_Yields_CI}) for the CI search and both channels are presented in Table~\ref{Yields_ADD} for the ADD LED search. Dilepton mass distributions for data and the predicted background are shown in Fig.~\ref{Mass} for both channels, along with a few benchmark CI and ADD signals overlaid.

\begin{table*}[Htbp]
\newcommand\T{\rule{0pt}{2.6ex}}
\newcommand\B{\rule[-1.2ex]{0pt}{0pt}}
\caption{Expected and observed event yields in the dielectron channel. The predicted yields are
shown for SM background as well as for SM+CI for several CI signal scenarios.
The quoted errors consist of both the statistical and systematic uncertainties added in quadrature.}
\scriptsize
\label{Electron_Yields_CI}
\centering
\vspace{0.2cm}
\begin{tabular}{c|c|c|c|c|c|c}
\hline
\hline
\multirow{2}{*}{Process} & \multicolumn{6}{c}{$m_{ee}$ [\GeV{}]} \T \B \\
\cline{2-7}
 & 400--550 \T & 550--800 & 800--1200 & 1200--1800 & 1800--3000 & 3000--4500 \\
\hline
Drell--Yan \T & 910 $\pm$ 70 & 302 $\pm$ 25 & 63 $\pm$ 6 & 8.2 $\pm$ 1.2 & 0.64 $\pm$ 0.17 & 0.006 $\pm$ 0.004 \\
Top quarks & 153 $\pm$ 13 & 35.2 $\pm$ 2.7 & 3.06 $\pm$ 0.18 & 0.140 $\pm$ 0.008 & $<$ 0.004   & $<$ 0.001   \\
Multi-Jet \& W+Jets & 88 $\pm$ 18 & 27 $\pm$ 5 & 5.8 $\pm$ 1.2 & 0.87 $\pm$ 0.17 & 0.11 $\pm$ 0.02 & 0.0058 $\pm$ 0.0012 \\
Diboson  & 62.2 $\pm$ 3.5 & 22.3 $\pm$ 1.3 & 5.4 $\pm$ 0.4 & 0.83 $\pm$ 0.05 & 0.075 $\pm$ 0.006 & $<$ 0.001   \\
Photon-Induced  \B & 40 $\pm$ 40 & 17 $\pm$ 17 & 4 $\pm$ 4 & 0.7 $\pm$ 0.7 & 0.08 $\pm$ 0.08 & 0.0016 $\pm$ 0.0016 \\
\hline
Total SM \T \B & 1260 $\pm$ 100 & 404 $\pm$ 35 & 82 $\pm$ 9 & 10.8 $\pm$ 1.6 & 0.91 $\pm$ 0.21 & 0.014 $\pm$ 0.005 \\
\hline
Data \T \B& 1262 & 388 & 84 & 7 & 0 & 0 \\
\hline
SM+CI ($\Lambda^{-}_{\rm {LL}} = 14 \TeV$) \T & 1310 $\pm$ 110 & 440 $\pm$ 40 & 108 $\pm$ 10 & 20.9 $\pm$ 1.9 & 4.2 $\pm$ 0.4 & 0.141 $\pm$ 0.028 \\
SM+CI ($\Lambda^{-}_{\rm {LL}} = 20 \TeV$) \T & 1290 $\pm$ 110 & 430 $\pm$ 40 & 90 $\pm$ 10 & 14.4 $\pm$ 1.7 & 2.01 $\pm$ 0.25 & 0.045 $\pm$ 0.012 \\
SM+CI ($\Lambda^{-}_{\rm {LR}} = 14 \TeV$) \T & 1340 $\pm$ 110 & 460 $\pm$ 40 & 118 $\pm$ 10 & 26.3 $\pm$ 2.1 & 6.0 $\pm$ 0.5 & 0.28 $\pm$ 0.05 \\
SM+CI ($\Lambda^{-}_{\rm {LR}} = 20 \TeV$) \T & 1290 $\pm$ 110 & 420 $\pm$ 40 & 98 $\pm$ 10 & 15.7 $\pm$ 1.7 & 2.58 $\pm$ 0.28 & 0.078 $\pm$ 0.018 \\
SM+CI ($\Lambda^{-}_{\rm {RR}} = 14 \TeV$) \T & 1310 $\pm$ 110 & 440 $\pm$ 40 & 108 $\pm$ 10 & 20.8 $\pm$ 1.9 & 3.78 $\pm$ 0.34 & 0.23 $\pm$ 0.04 \\
SM+CI ($\Lambda^{-}_{\rm {RR}} = 20 \TeV$) \T \B & 1290 $\pm$ 110 & 430 $\pm$ 40 & 91 $\pm$ 10 & 14.3 $\pm$ 1.7 & 1.86 $\pm$ 0.24 & 0.072 $\pm$ 0.015 \\
\hline
SM+CI ($\Lambda^{+}_{\rm {LL}} = 14 \TeV$) \T & 1230 $\pm$ 110 & 380 $\pm$ 40 & 79 $\pm$ 9 & 12.2 $\pm$ 1.7 & 2.08 $\pm$ 0.25 & 0.075 $\pm$ 0.015 \\
SM+CI ($\Lambda^{+}_{\rm {LL}} = 20 \TeV$) \T & 1230 $\pm$ 110 & 390 $\pm$ 40 & 77 $\pm$ 9 & 10.0 $\pm$ 1.6 & 0.95 $\pm$ 0.22 & 0.029 $\pm$ 0.008 \\
SM+CI ($\Lambda^{+}_{\rm {LR}} = 14 \TeV$) \T & 1200 $\pm$ 110 & 400 $\pm$ 40 & 88 $\pm$ 10 & 18.9 $\pm$ 1.8 & 4.2 $\pm$ 0.4 & 0.191 $\pm$ 0.034 \\
SM+CI ($\Lambda^{+}_{\rm {LR}} = 20 \TeV$) \T & 1210 $\pm$ 110 & 390 $\pm$ 40 & 81 $\pm$ 9 & 11.5 $\pm$ 1.6 & 1.65 $\pm$ 0.24 & 0.058 $\pm$ 0.013 \\
SM+CI ($\Lambda^{+}_{\rm {RR}} = 14 \TeV$) \T & 1230 $\pm$ 110 & 380 $\pm$ 40 & 79 $\pm$ 9 & 12.1 $\pm$ 1.7 & 2.26 $\pm$ 0.26 & 0.098 $\pm$ 0.018 \\
SM+CI ($\Lambda^{+}_{\rm {RR}} = 20 \TeV$) \T \B & 1230 $\pm$ 110 & 390 $\pm$ 40 & 77 $\pm$ 9 & 10.2 $\pm$ 1.6 & 1.06 $\pm$ 0.22 & 0.036 $\pm$ 0.009 \\
\hline
\hline
\end{tabular}
\end{table*}

\begin{table*}[Htbp]
\newcommand\T{\rule{0pt}{2.6ex}}
\newcommand\B{\rule[-1.2ex]{0pt}{0pt}}
\caption{Expected and observed event yields in the dimuon channel. The predicted yields are
shown for SM background as well as for SM+CI for several CI signal scenarios. 
The quoted errors consist of both the statistical and systematic uncertainties added in quadrature.}
\scriptsize
\label{Muon_Yields_CI}
\centering
\vspace{0.2cm}
\begin{tabular}{c|c|c|c|c|c|c}
\hline
\hline
\multirow{2}{*}{Process} & \multicolumn{6}{c}{$m_{\mu\mu}$ [\GeV{}]} \T \B \\
\cline{2-7}
 & 400--550 \T & 550--800 & 800--1200 & 1200--1800 & 1800--3000 & 3000--4500 \\
 \hline
Drell--Yan \T            & $670 \pm 50$           & $217 \pm 18$              & $45 \pm 5$               & $5.9 \pm 0.8$               & $0.58 \pm 0.12$                  & $0.027 \pm 0.008$  \\
Top quarks                           & $128 \pm 10$            & $16.3 \pm 1.4$           & $1.66 \pm 0.11$      & $0.103 \pm 0.007$            & $ < 0.005 $                     & $ < 0.002 $ \\
Diboson                & $47.6 \pm 2.7$          & $15.3 \pm 0.9$           & $3.75 \pm 0.26$      & $0.556 \pm 0.030$      & $0.056 \pm 0.005$              & $ < 0.003 $ \\
Photon-Induced \B & $34 \pm 34$               & $13 \pm 13$               & $3.3 \pm 3.3$           & $0.5 \pm 0.5$               & $0.07 \pm 0.07$                  & $ < 0.006 $  \\
\hline 
Total SM     \T \B              &$880 \pm 60$             & $261 \pm 22$               & $54 \pm 6$                & $7.2 \pm 1.0 $             & $0.71 \pm 0.14$                 & $0.032 \pm 0.009$ \\
\hline 
Data          \T \B                                & $814$                 & $265$                          & $47$                         & $7$                                 & $1$                                          & $0$
 \\
\hline     
SM+CI ($\Lambda^{-}_{\rm {LL}} = 14 \TeV$) \T & $900 \pm 60$        & $285 \pm 23$         & $70 \pm 6$         & $14.4 \pm 1.2$         & $2.89 \pm 0.33$         & $0.18 \pm 0.04$ \\
SM+CI ($\Lambda^{-}_{\rm {LL}} = 20 \TeV$) \T & $870 \pm 60$         & $265 \pm 23$         & $58 \pm 6$         & $10.0 \pm 1.1$         & $1.49 \pm 0.18$         & $0.103 \pm 0.022$ \\    
SM+CI ($\Lambda^{-}_{\rm {LR}} = 14 \TeV$) \T & $930 \pm 60$         & $292 \pm 23$         & $79 \pm 6$         & $16.9 \pm 1.4$         & $3.9 \pm 0.4$            & $0.38 \pm 0.08$ \\
SM+CI ($\Lambda^{-}_{\rm {LR}} = 20 \TeV$) \T & $910 \pm 60$         & $281 \pm 23$         & $61 \pm 6$         & $10.7 \pm 1.1$         & $1.76 \pm 0.20$         & $0.139 \pm 0.029$ \\
SM+CI ($\Lambda^{-}_{\rm {RR}} = 14 \TeV$) \T & $900 \pm 60$         & $285 \pm 23$         & $70 \pm 6$         & $13.8 \pm 1.2$         & $2.80 \pm 0.32$         & $0.20 \pm 0.04$\\
SM+CI ($\Lambda^{-}_{\rm {RR}} = 20 \TeV$) \T \B & $870 \pm 60$         & $265 \pm 23$         & $58 \pm 6$         & $10.1 \pm 1.1$        & $1.29 \pm 0.17$         & $0.09 \pm 0.02$ \\
\hline       
SM+CI ($\Lambda^{+}_{\rm {LL}} = 14 \TeV$) \T  & $870 \pm 60$         & $252 \pm 23$         & $51 \pm 6$         & $7.5 \pm 1.0$         & $1.45 \pm 0.18$         & $0.113 \pm 0.023$ \\
SM+CI ($\Lambda^{+}_{\rm {LL}} = 20 \TeV$) \T & $890 \pm 60$         & $247 \pm 23$         & $50 \pm 6$          & $6.4 \pm 1.0$         & $0.74 \pm 0.15$         & $0.048 \pm 0.013$ \\
SM+CI ($\Lambda^{+}_{\rm {LR}} = 14 \TeV$) \T & $860 \pm 60$         & $256 \pm 23$         & $57 \pm 6$         & $12.2 \pm 1.1$         & $2.79 \pm 0.31$         & $0.28 \pm 0.06$ \\
SM+CI ($\Lambda^{+}_{\rm {LR}} = 20 \TeV$) \T & $880 \pm 60$         & $252 \pm 23$         & $50 \pm 6$         & $7.5 \pm 1.0$         & $1.15 \pm 0.16$         & $0.092 \pm 0.019$ \\
SM+CI ($\Lambda^{+}_{\rm {RR}} = 14 \TeV$) \T & $870 \pm 60$         & $252 \pm 23$         & $51 \pm 6$         & $8.0 \pm 1.0$                 & $1.36 \pm 0.18$         & $0.138 \pm 0.026$ \\
SM+CI ($\Lambda^{+}_{\rm {RR}} = 20 \TeV$) \T \B & $890 \pm 60$         & $247 \pm 23$         & $50 \pm 6$         & $6.5 \pm 1.0$         & $0.70 \pm 0.15$         & $0.052 \pm 0.013$\\
\hline
\hline
\end{tabular}
\end{table*}

\begin{figure}[tbp]
 \begin{center}
  \includegraphics[width=0.49\textwidth]{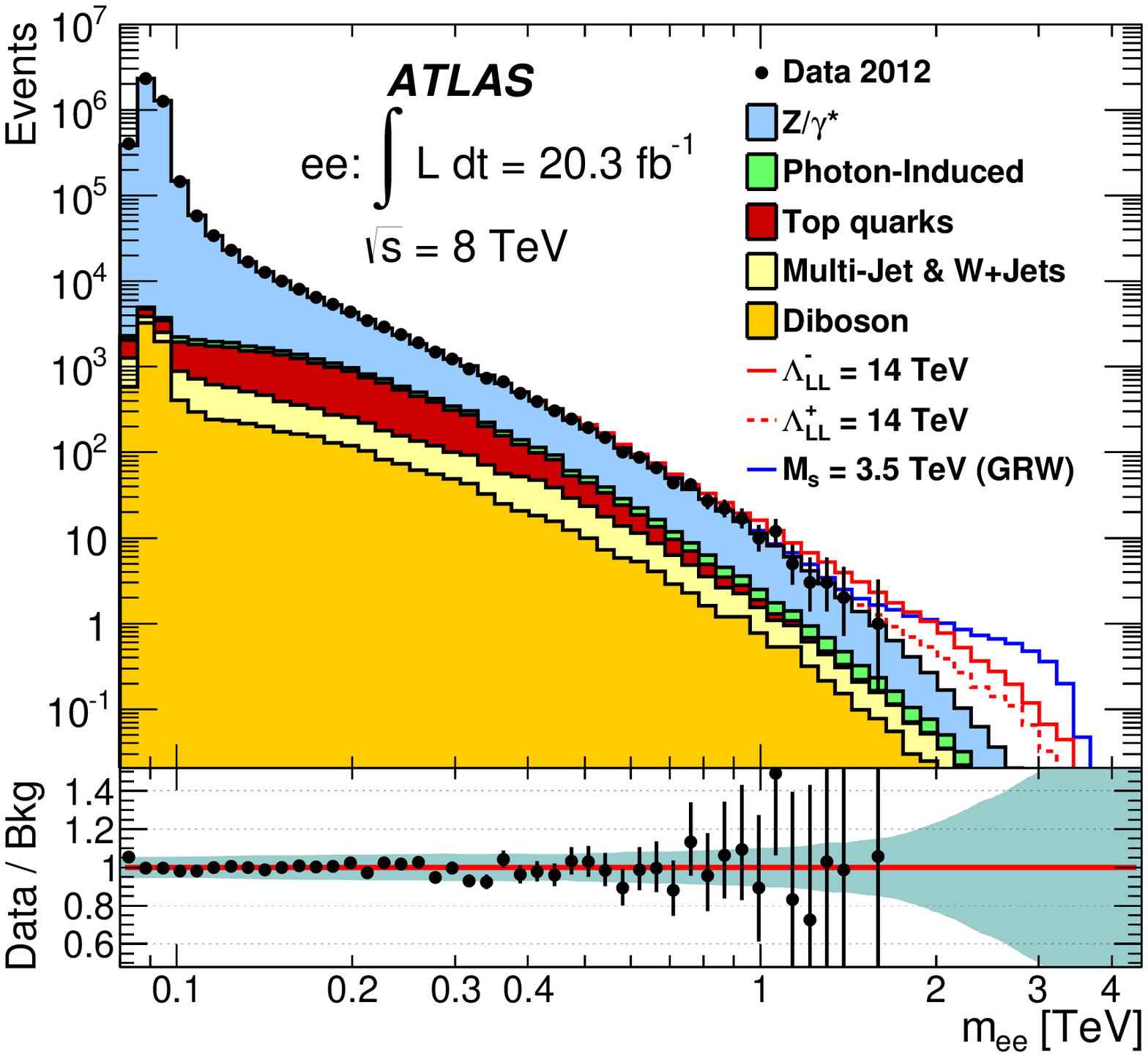}
  \includegraphics[width=0.49\textwidth]{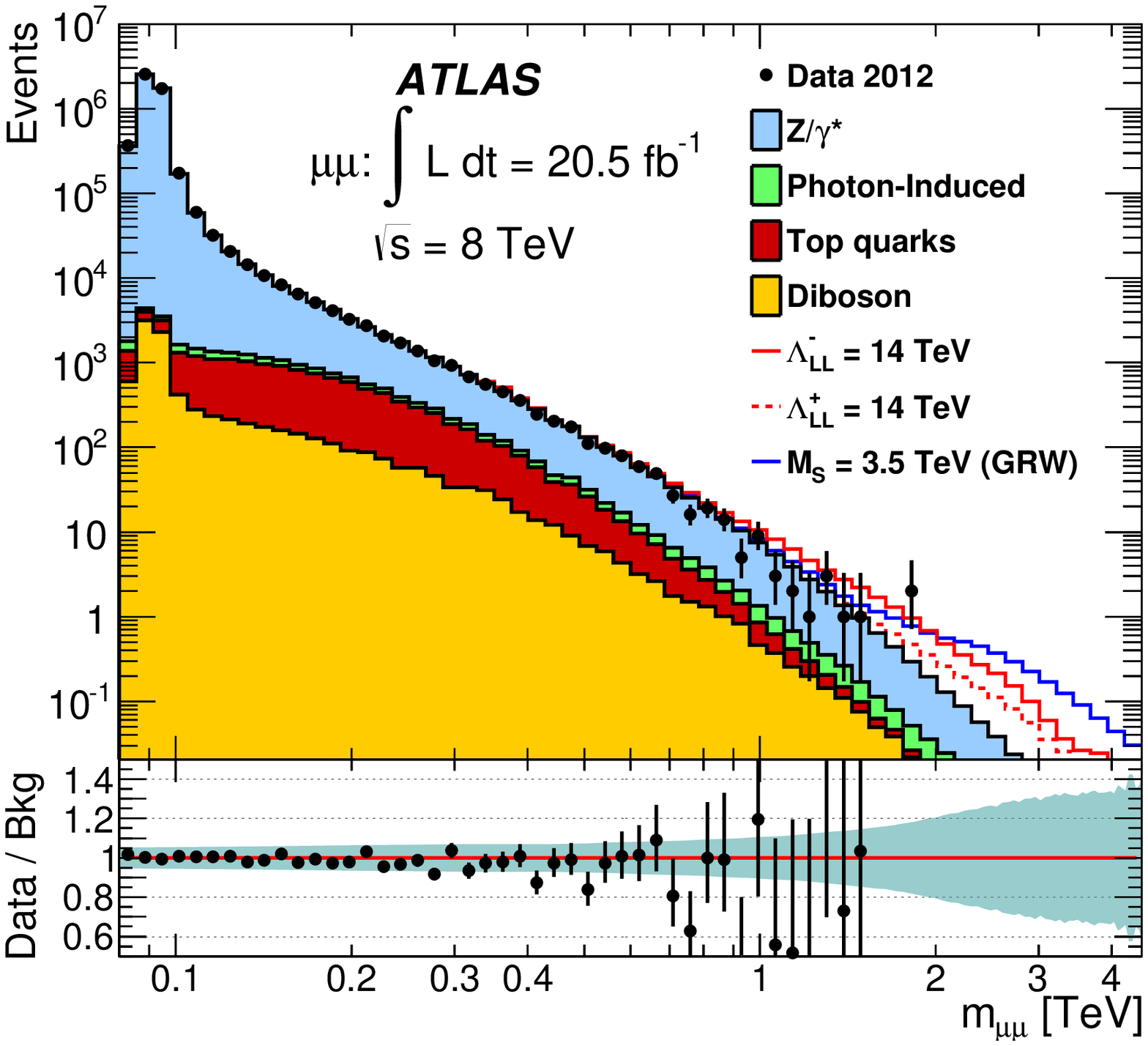}
 \end{center}
 \caption{Reconstructed dielectron (top) and dimuon (bottom) mass distributions for data and the SM background estimate. Also shown are the predictions for a benchmark $\Lambda$ value in the LL contact interaction model and benchmark $M_{\rm S}$ value in the GRW ADD model. The distribution bin width is constant in $\log(m_{\ell\ell})$. The ratio is presented with the total systematic uncertainty overlaid as a band.}
 \label{Mass}
\end{figure}

The dilepton invariant mass is commonly used as the discriminating variable for a CI search.
However, the lepton decay angle also has high discriminating power from DY events in certain cases such as the left-right model. Therefore, the dilepton decay angle, $\theta^*$, is also used as a discriminating variable in the CI search.
The angle $\theta^\ast$ is defined in the Collins--Soper (CS) frame~\cite{Collins:1977iv}, which is constructed with the $z$-axis bisecting the angle between the two incoming parton momenta, and the $x$-axis perpendicular to the incoming parton momentum plane. As the incoming parton information from $pp$ collisions is unknown, the direction of the dilepton system is taken to be the direction of the incoming quark (as opposed to anti-quark). This introduces a dilution of any asymmetry in the $\cos\theta^*$ distribution (leading to derived angular variables being described as ``uncorrected''). The angle $\theta^*$ is then taken as the angle between this $z$-axis and the outgoing negatively charged lepton, using the formula

\vspace{-0.5cm}

\begin{equation}\label{eq:CosThetaStar}
\footnotesize\cos\theta^* = \frac{p_z(\ell^+\ell^-)}{|p_z(\ell^+\ell^-)|} \frac{2(p_1^+p_2^- - p_1^-p_2^+)}{m(\ell^+\ell^-) \sqrt{m(\ell^+\ell^-)^2 + p_T(\ell^+\ell^-)^2}}~,\nonumber
\end{equation}

\noindent where $p_n^{\pm}$ denotes $\frac{1}{\sqrt{2}}(E \pm p_z)$ and $n$ = $1$ or $2$ corresponds to the negatively charged or positively charged leptons, respectively. From this angle, a forward-backward asymmetry, which is sensitive to the chiral structure of the interaction, is defined as follows:

\vspace{-0.3cm}

\begin{equation}\label{A_FB}
A_{\rm FB} = \frac{N_{\rm F} - N_{\rm B}}{N_{\rm F} + N_{\rm B}}~,\nonumber
\end{equation}

\noindent where $N_{\rm F}$ ($N_{\rm B}$) is the number of events with $\cos\theta^*$ greater (smaller) than zero. The discrimination between CI+SM and the SM-only background is due to the couplings of the CI model, which predicts a larger $A_{\rm {FB}}$ than the SM background for the CI signal in the left-left and right-right model, and an equally large but opposite-sign $A_{\rm {FB}}$ for the left-right model. If a CI signal were present in nature this would therefore lead to a modest increase in the total measured $A_{\rm {FB}}$ as a function of dilepton mass for the left-left and right-right model, and a substantial decrease in the measured $A_{\rm {FB}}$ for the left-right model. Therefore in the CI search, each dilepton mass bin is further divided into forward and backward events for the statistical interpretation of the results. Figures~\ref{CosThetaStar} and \ref{Afb} present the data and background for $\cos\theta^*$ and $A_{\rm {FB}}$ as a function of dilepton mass, respectively, in both channels. These distributions also display CI signal predictions.

Good agreement is observed between the data and the background model in both the dilepton mass and $A_{\rm FB}$ distributions.

\begin{table}[htbp]
\newcommand\T{\rule{0pt}{2.6ex}}
\newcommand\B{\rule[-1.2ex]{0pt}{0pt}}
\caption{Expected and observed event yields in the dielectron (second column) and dimuon (third column) channels in the ADD search for large extra dimensions. The expected yields for the SM plus two GRW ADD parameter points are also shown. The quoted errors consist of both the statistical and systematic uncertainties added in quadrature.}
\footnotesize
\label{Yields_ADD}
\centering
\vspace{0.2cm}
\begin{tabular}{c|c|c}
\hline
\hline
\multirow{2}{*}{Process} & $m_{ee}$ [\GeV{}] \T \B & $m_{\mu\mu}$ [\GeV{}] \\
\cline{2-3}
 & 1900--4500 \T & 1900--4500 \\
\hline
Drell--Yan \T           & 0.43 $\pm$ 0.12 & 0.44 $\pm$ 0.09 \\
Top quarks \T                 & $<$ 0.002 & $<$ 0.006 \\
Multi-Jet \& $W$+Jets \T & 0.062 $\pm$ 0.012 & $<$ 0.001 \\
Diboson \T             & 0.053 $\pm$ 0.005 & 0.047 $\pm$ 0.005 \\
Photon-Induced \T\B    & 0.06  $\pm$ 0.06 & 0.05 $\pm$ 0.05 \\
\hline
Total SM \T \B         & 0.61 $\pm$ 0.13 & 0.54 $\pm$ 0.09 \\
\hline
Data \T \B             & 0 & 0 \\
\hline
SM+ADD ($M_{\rm S} = 3.5 \TeV$) \T    & 5.8  $\pm$ 0.5  & 3.9  $\pm$ 0.4 \\
SM+ADD ($M_{\rm S} = 4.0 \TeV$) \T \B & 2.56 $\pm$ 0.24 & 1.69 $\pm$ 0.14 \\
\hline
\hline
\end{tabular}
\end{table}

\begin{figure}[Htbp]
 \begin{center}
  \includegraphics[width=0.49\textwidth]{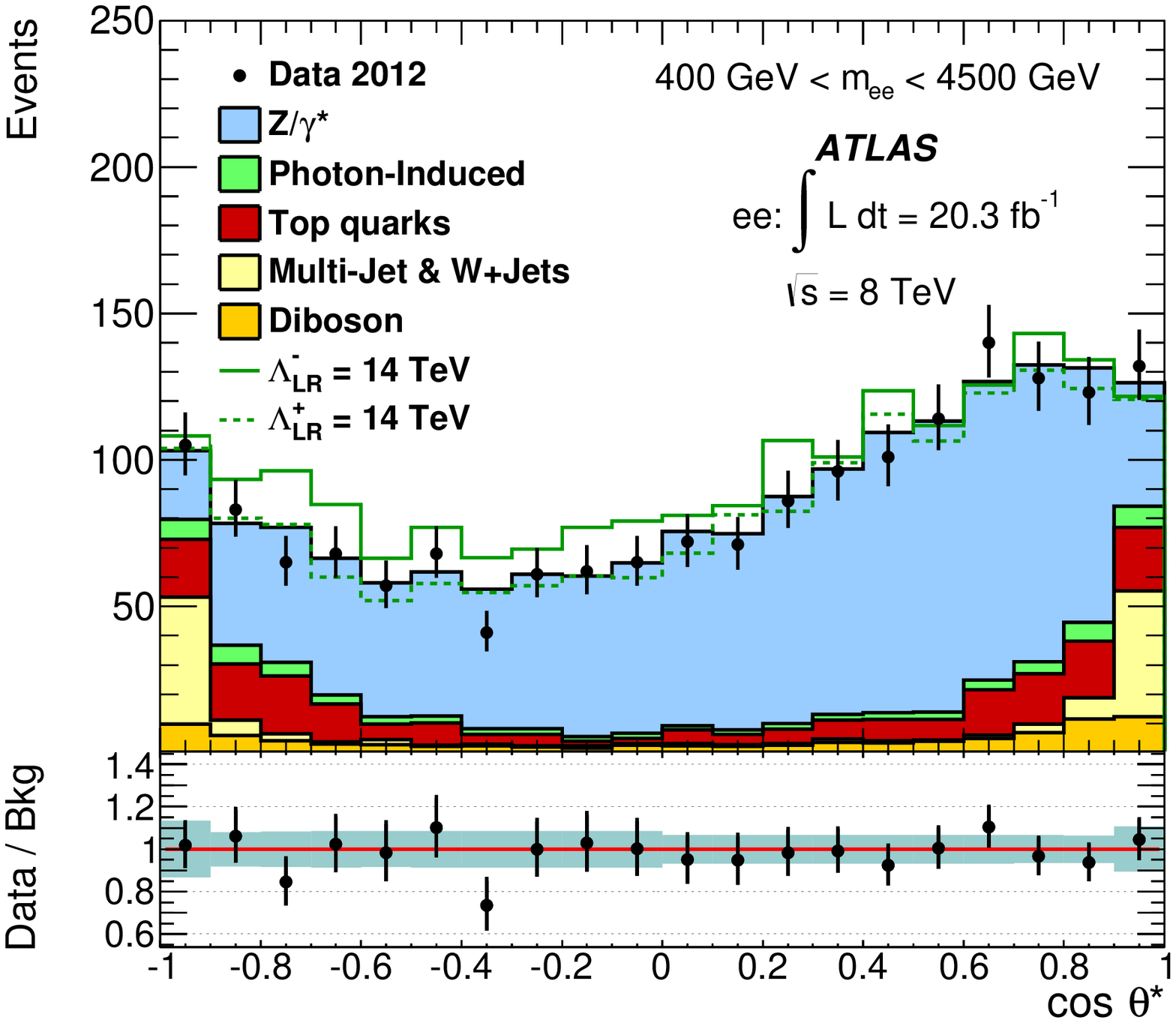}
  \includegraphics[width=0.49\textwidth]{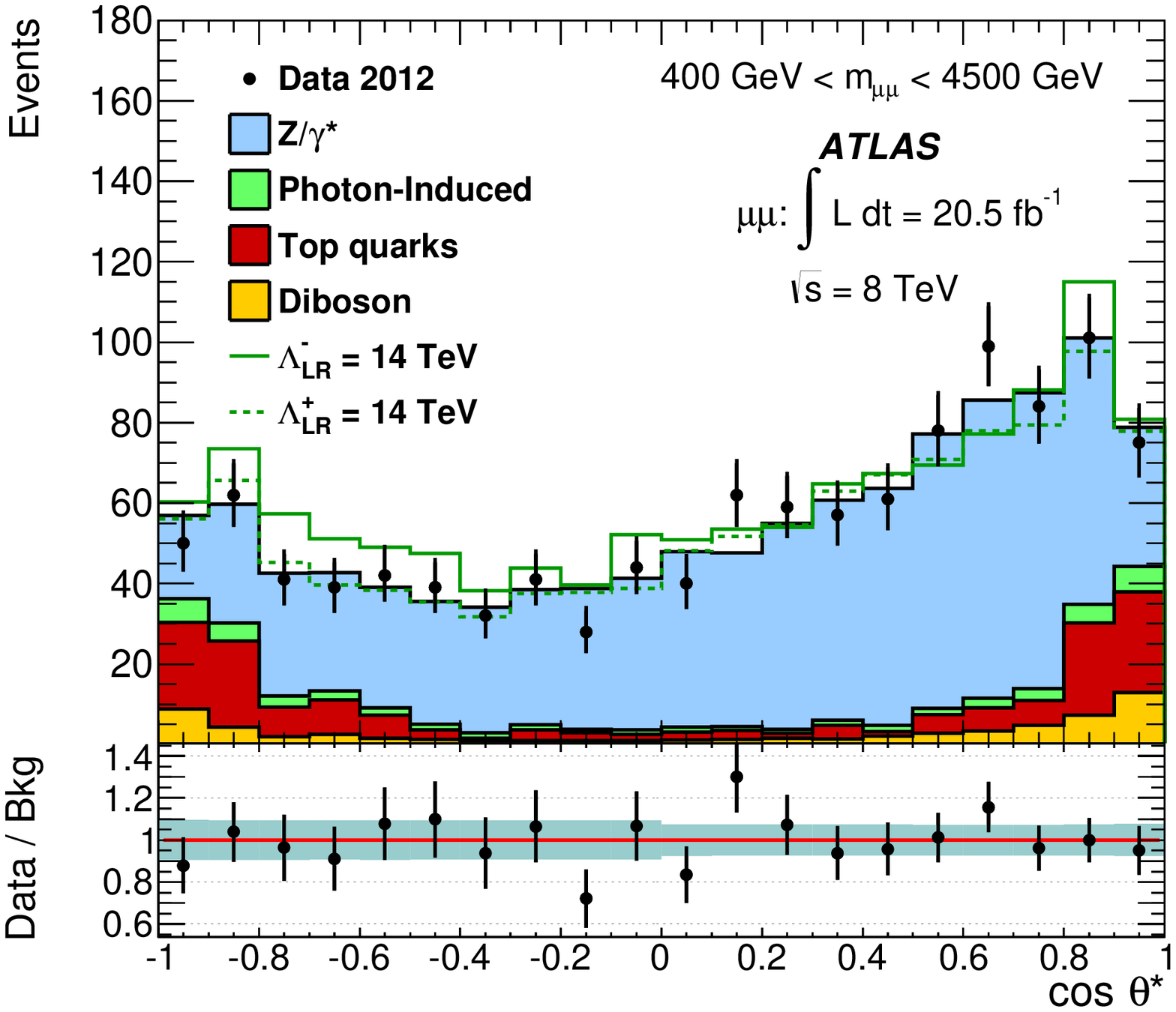}
 \end{center}
 \caption{Reconstructed $\cos\theta^*$ distributions for data and the SM background estimate in the dielectron (top) and dimuon (bottom) channels.  Results are shown for the contact interaction signal region for dilepton masses between 400~\GeV{} and 4500~\GeV{}. Also shown are the predictions for a benchmark $\Lambda$ value in the LR contact interaction model. The ratio is presented with the total systematic uncertainty overlaid as a band.}
 \label{CosThetaStar}
\end{figure}

\begin{figure}[Htbp]
 \begin{center}
  \includegraphics[width=0.49\textwidth]{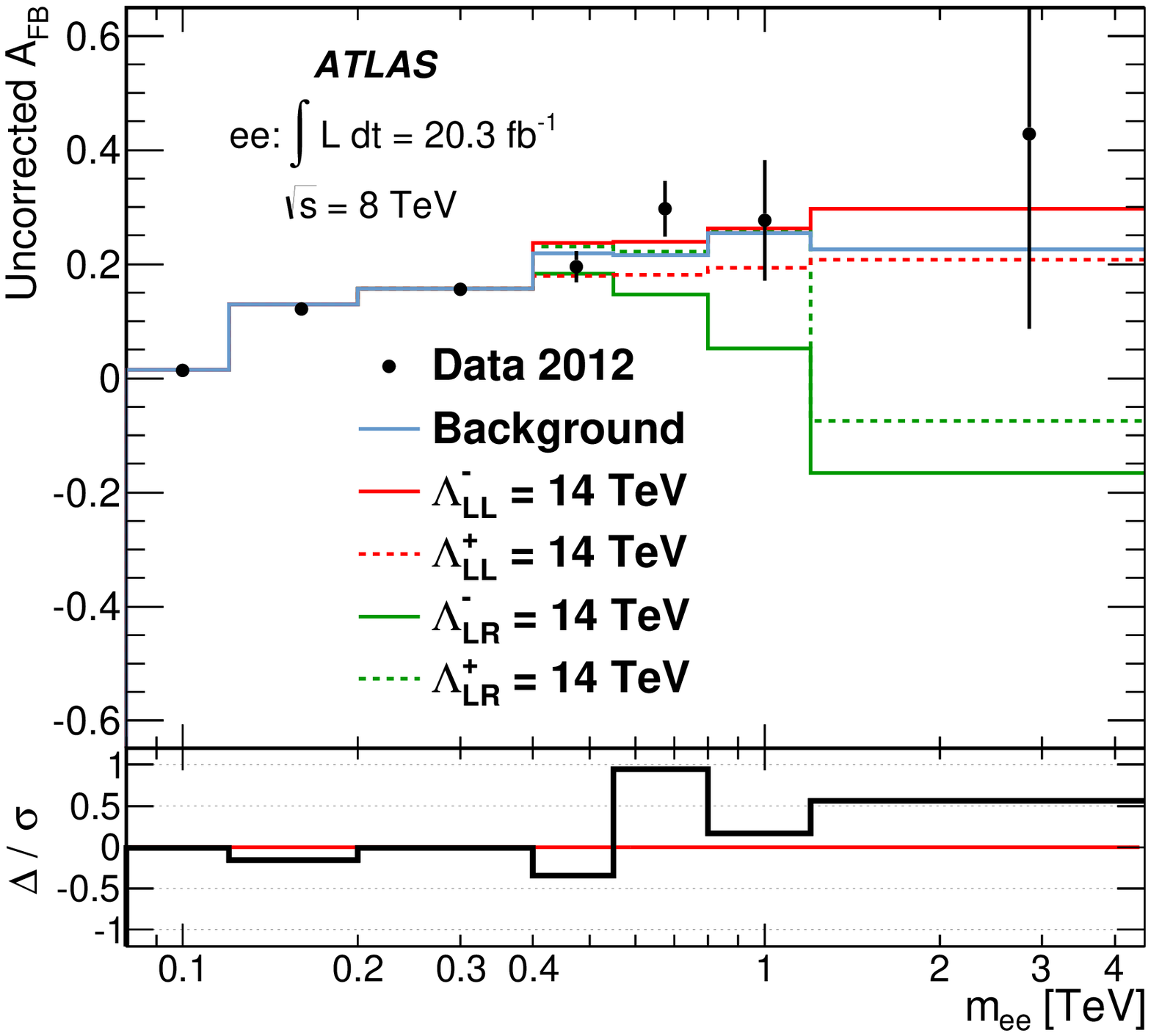}
  \includegraphics[width=0.49\textwidth]{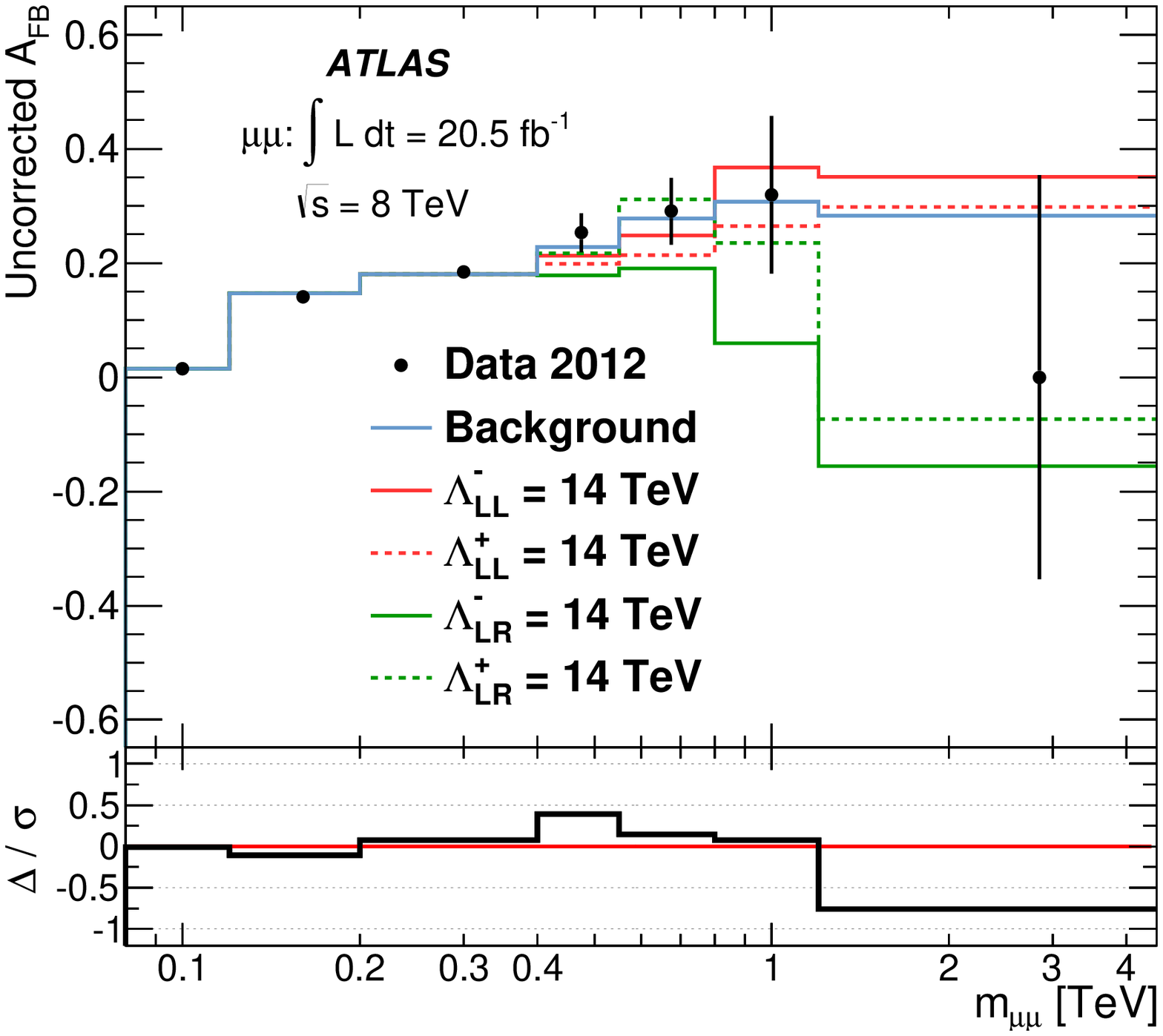}
 \end{center}
 \caption{Reconstructed $A_{\rm FB}$ distributions for data and the SM background estimate as a function of dielectron (top) and dimuon (bottom) mass. Also shown are the predictions of different benchmark $\Lambda$ values for the LL and LR contact interaction model (the RR model is very similar to the LL case). The ratio displays the background-subtracted data ($\Delta$) divided by the total uncertainty ($\sigma$) in each bin.}
 \label{Afb}
\end{figure}


\section{Systematic uncertainties}\label{SectionSystematics}

The total background estimate is normalized by scaling to data in the dilepton mass region 80--120~\GeV{}. This protects the analysis against mass-independent systematic uncertainties (such as the luminosity uncertainty), as any constant scale factor cancels. However, mass-dependent systematic uncertainties affect the shape of the discriminating variables and are therefore considered as nuisance parameters in the statistical interpretation.

Experimental uncertainties originate from the following sources: lepton trigger and reconstruction efficiencies, lepton energy and momentum scale and resolution, lepton charge misidentification, multi-jet \& $W$+jets background estimate (in the $ee$ channel), beam energy scale, and MC statistics.

It is important to control the lepton momentum uncertainty, as mismodelling of the resolution could result in a broad signal-like excess (or deficit) in the dilepton mass distribution. The muon momentum resolution depends critically on the quality of the MS chamber alignment. Resolution uncertainties are determined from dedicated data-taking periods with no magnetic field in the MS and from studies of muon tracks passing through the overlap region between chambers in the small and large sectors of the MS where the independent track momenta reconstructed from these adjacent sectors can be compared directly. The electron momentum uncertainty is negligible.

Another important experimental uncertainty is the charge misidentification which can arise from two main sources: track curvature and ``trident'' events. The latter occurs when a hard bremsstrahlung is emitted by a high-momentum lepton, and a subsequent photon conversion gives rise to a high-momentum track with a charge opposite to that of the initial lepton, but which is selected erroneously. To study the trident source of charge misidentification, dedicated MC samples were generated with the amount of detector material varied by up to 20\% of a radiation length. To study the track curvature source, various investigations were carried out wherein additional charge misidentification is injected into the simulation to ascertain its effect, and the ID track resolution in $q/p$ is varied to assess the probability of a change of charge sign. As these systematic uncertainty studies found a negligible change in charge misidentification, a conservative uncertainty of 20\% with respect to the measured charge misidentification rate in Drell--Yan MC simulation was applied. 
For the dielectron channel, the charge misidentification systematic uncertainty can be as large as 3\%. For the dimuon channel, this is covered by the resolution uncertainty and is negligible.

The uncertainty on the data-driven estimate of the combined multi-jet \& $W$+jets background is assessed by comparing complementary estimation methods (giving a maximum deviation of 18\% from the nominal method) and variations of the real-electron contamination suppression requirements in the nominal method (resulting in deviations of up to 5\%). The addition of these effects in quadrature gives a total systematic uncertainty on the data-driven estimate of 20\%.  A detailed description of this procedure is given in Ref.~\cite{Aad:Zprime}.

A systematic uncertainty on the LHC beam energy of 0.65\%~\cite{Wenninger:1546734} is assessed for both the signal and background processes.

The statistical uncertainty of the MC samples is included as a systematic uncertainty for both the signal and the background.  This includes the fit uncertainty due to the high-mass extrapolation of the top-quarks background, which is described in Sect.~\ref{SectionSelection}.

The theoretical uncertainties are the variations among the PDF eigenvector sets, the effect of PDF choice, the PDF $\alpha_S$ scale, the EW higher-order corrections, the photon-induced contributions, and the DY cross-section uncertainty. 
The effect of these uncertainties on the background yield are taken into account with a standard procedure where event weights are used to create systematically shifted distributions, which are then used as nuisance parameters in the statistical interpretation.  However, for the signal yields one does not want to introduce a bias via the specific theoretical uncertainty choices, and therefore these are only taken into account by the effect that they have on the signal acceptance times efficiency. This effect was found to be negligible in all cases except for the PDF variation in the ADD search, where an additional uncertainty of 6\% (3\%) is included in the dielectron (dimuon) channel as a nuisance parameter in the statistical interpretation.
For the CI search, systematic uncertainties are taken into account as a function of dilepton mass for forward and backward events separately, to account for any variation in the uncertainty which might affect the expected asymmetry.  For example, the largest systematic uncertainty in this search is the background PDF variation, which has an effect in the dielectron (dimuon) channel of 11\% (12\%) at a mass of 2~\TeV{}.  When separated into forward and backward regions at the same dielectron (dimuon) mass, this uncertainty is 10\% (8.5\%) and 16\% (15\%), respectively.  Likewise the PI uncertainty in the dielectron (dimuon) channel of 12\% (9.5\%) at a mass of 2~\TeV{}, becomes 10\% (7.5\%) and 16\% (13\%), when separated into forward and backward regions, respectively. The other sources of systematic uncertainty were found to not have a strong dependence between forward and backward events. 
The different sources of PDF uncertainty are assessed by utilizing the 
MSTW2008NNLO PDF error set (90\% C.L.) and by following the procedure detailed 
in Ref.~\cite{Aad:Zprime,Botje:2011sn}. The uncertainty due to the choice of PDF is investigated by comparing the central values of various PDFs, namely MSTW2008NNLO, CT10NNLO~\cite{Gao:2013xoa}, NNPDF2.3~\cite{Ball:2012cx}, ABM11~\cite{Alekhin:2013dmy}, and HERAPDF1.5~\cite{CooperSarkar:2011aa}.  All except for ABM11 are found to be within the MSTW2008NNLO 90\% C.L. uncertainty, and so the variation from ABM11 with respect to the MSTW2008NNLO central value, outside of the MSTW2008NNLO 90\% C.L. uncertainty, is taken as a separate systematic uncertainty due to PDF choice. VRAP~\cite{Anastasiou:2003ds} is used to assess the $\alpha_S$ systematic uncertainty, along with scale uncertainties which are estimated by varying the nominal renormalization and factorization scales simultaneously by a factor of two.
A study to ascertain the photon-induced background estimate uncertainty was performed in Ref.~\cite{Aad:Zprime}, and found that the nominal MRST2004QED PDF gives an upper estimate of the PI contribution. Varying the assumed quark masses showed that the lower bound of this estimate gives fairly small PI contributions. Therefore the PI background estimate is assigned a conservative uncertainty of 100\%. 
A uniform uncertainty of 4\% due to the uncertainty on the Z/$\gamma^*$ NNLO cross-section (using MSTW2008NNLO 90\% C.L.) in the normalization region was determined in Ref.~\cite{Aad:Zprime} and is applied to signal event yields due to the normalization procedure. The variation due to the cross-section uncertainty in the other background MC samples was found to be negligible.
All systematic uncertainties are treated as uncorrelated, and a summary of the systematic uncertainties at dilepton masses of 1 and 2~\TeV{} is presented in Table~\ref{systematics}.

\begin{table*}[htbp]
\newcommand\T{\rule{0pt}{2.6ex}}
\newcommand\B{\rule[-1.2ex]{0pt}{0pt}}
  \begin{center}
    \caption{Summary of the systematic uncertainties taken into account for the total expected number of events. Values are provided at $m_{\ell\ell}$ = 1~\TeV{} (2~\TeV{}) to give representative estimates relevant to this search. The PDF variation values shown for signal are based on CI. For the ADD signal they are uniform at 6\% and 3\% in the dielectron and dimuon channels, respectively. Signal systematic uncertainties are assessed as a function of the corresponding parameter of interest but are not found to vary greatly. N/A indicates that the uncertainty is not applicable.
    \label{systematics}}
    \centering
    \addtolength{\tabcolsep}{+1pt}
    \vspace{0.2cm}
    \begin{tabular}{c|cc|cc}
      \hline
      \hline
      \multirow{2}{*}{Source} & \multicolumn{2}{c|}{Dielectrons \T} & \multicolumn{2}{c}{Dimuons} \\
      \cline{2-5}
                        & Signal \T          & Background       & Signal                 & Background    \\
      \hline
      Normalization \T  & 4.0\% ~(4.0\%)     & N/A               & 4.0\% ~(4.0\%)         & N/A \\
      PDF Variation     & $<$ 0.1\% ~(0.2\%) & 5.0\% ~(11.0\%)  &  $<$ 0.1\% ~($<$ 0.1\%)& 5.0\% ~(12.0\%) \\
      PDF Choice        & N/A                 & 1.0\% ~(7.0\%)   & N/A                     & 1.0\% ~(6.0\%) \\
      $\alpha_S$        & N/A                 & 1.0\% ~(3.0\%)   & N/A                     & 1.0\% ~(3.0\%) \\
      EW Corrections    & N/A                 & 1.0\% ~(2.0\%)   & N/A                     & 1.0\% ~(3.0\%) \\
      Photon-Induced    & N/A                 & 7.0\% ~(12.0\%)  & N/A                     & 6.5\% ~(9.5\%)  \\
      Efficiency        & 1.0\% ~(2.0\%)     & 1.0\% ~(2.0\%)   & 3.0\% ~(6.0\%)         & 3.0\% ~(6.0\%) \\
      Scale \& Resolution  & 1.2\% ~(2.4\%)     & 1.2\% ~(2.4\%)   & 1.0\% ~(4.0\%)         & 1.0\% ~(4.0\%) \\
      Electron Charge Misident. & 1.2\% ~(2.0\%)     & 1.2\% ~(2.0\%)   & N/A & N/A  \\
      Multi-Jet \& $W$+Jets& N/A                 & 3.0\% ~(5.0\%)   & N/A                     & N/A \\
      Beam Energy       & 1.0\% ~(3.0\%)     & 1.0\% ~(3.0\%)   & 1.0\% ~(3.0\%)         & 2.0\% ~(3.0\%)  \\
      MC Statistics   \B  & 3.0\% ~(3.0\%)     & 0.5\% ~(0.5\%)   & 3.0\% ~(3.0\%)         & 0.5\% ~(0.5\%)  \\
      \hline
      Total       \T\B  & 5.5\% ~(6.9\%)     & 9.5\% ~(19.4\%) & 6.0\% ~(9.3\%)         & 9.2\% ~(18.7\%) \\
      \hline
      \hline
    \end{tabular}
  \end{center}
\end{table*}


\section{Statistical interpretation}

A Bayesian approach is used for the statistical interpretation of the results, using a uniform positive prior as a function of the parameter of interest to quantify any observed excess. In the absence of a signal, 95\% C.L. lower exclusion limits are set on that parameter.
The total number of expected events $\mu$ in each search region can be expressed as

\vspace{-0.3cm}

\begin{equation}\label{eq:mu}
\mu = n_{\rm s}(\Theta,\overline{\Omega}) + n_{\rm b}(\overline{\Omega}),\nonumber
\end{equation}

\noindent where $n_{s}(\Theta,\overline{\Omega})$ is the number of events predicted by the CI or ADD signal for a particular choice of model parameter $\Theta$. The quantity $n_{b}(\overline{\Omega})$ is the total number of background events, and in both cases $\overline{\Omega}$ represents the set of Gaussian nuisance parameters that account for systematic uncertainties on the number of respective signal and background events. The parameter $\Theta$ corresponds to a choice of contact interaction scale $\Lambda$ and interference parameter $\eta_{ij}$ in the case of the CI interpretation, and a choice of string scale $M_{\rm S}$ and specific formalism (GRW, Hewett, or HLZ) in the case of the ADD interpretation.

The likelihood of observing $n$ events given the new physics parameter $\Theta$ and nuisance parameters $\overline{\Omega}$ is then the product of Poisson probabilities for each mass--$\cos\theta^*$ bin $k$:

\vspace{-0.3cm}

\begin{equation}\label{likelihood}
 \text{$\cal L$} (n \mid \Theta,\overline{\Omega}) = \displaystyle\prod_{l=1}^{N_{\rm {channel}}}\displaystyle\prod_{k=1}^{N_{\rm {bin}}}\frac{\mu_{lk}^{n_{lk}}e^{-\mu_{lk}}}{n_{lk}!},\nonumber
\end{equation}

\noindent where $n_{lk}$ is the number of events observed in data, and $\mu_{lk}$ is the total number of expected events (signal plus background), both in mass--$\cos\theta^*$ bin $k$ and channel $l$ (where the channel can be dielectron or dimuon). According to Bayes' theorem, the posterior probability for the parameter $\Theta$, given $n$ observed events, is then

\vspace{-0.3cm}

\begin{equation}\label{posterior}
\text{$\cal P$} (\Theta \mid n) = \frac{1}{\text{$\cal Z$}} \text{$\cal L_M$} (n \mid \Theta) P(\Theta),\nonumber
\end{equation}

\noindent where ${\cal Z}$ is a normalization constant and the marginalized likelihood $\cal L_M$ corresponds to the likelihood after all nuisance parameters are integrated out. This integration is performed assuming that the nuisance parameters are correlated across all dilepton mass--$\cos\theta^*$ bins. The nuisance parameters that are treated as correlated between both channels are: PDF uncertainties, EW corrections, photon-induced, beam energy, and normalization. All other sources are treated as uncorrelated.
Table~\ref{systematics} shows which nuisance parameters are taken into account for the signal and background expectations. The prior probability $P(\Theta)$ is chosen to be uniform and positive in either $1/\Lambda^2$ or $1/\Lambda^4$ for the CI analysis, and either $1/M_{\rm S}^4$ or $1/M_{\rm S}^8$ for the ADD analysis. These choices are motivated by the form of Eqs.~(\ref{cross_section_CI}) and (\ref{cross_section_ADD}), to give the reader a sense of how the interplay in these forms can affect the result. The 95\% C.L. limit is then obtained by finding the value $\Theta_{\rm lim}$ satisfying $\int_0^{\Theta_{\rm lim}} \text{$\cal P$} ($$\Theta$$\mid n) \: d\Theta = 0.95$, where $\Theta$ is chosen to be $1/\Lambda^2$, $1/\Lambda^4$, $1/M_{\rm S}^4$ or $1/M_{\rm S}^8$.

The calculations are performed with the Bayesian Analysis Toolkit~\cite{Caldwell:2008fw}, which uses a Markov Chain Monte Carlo technique to integrate over the nuisance parameters.  For each physics model, 1000 pseudo-experiments (PEs) are run to obtain an adequate SM-only expected distribution; the PE with the median parameter of interest value provides the expected limit, with $\pm 1\:\sigma$  and $\pm 2\:\sigma$ intervals also obtained from this set of 1000 PEs correspondingly.  In order to quantify the consistency between the data and the background expectation, the likelihood ratio is computed for the signal-plus-background and background-only hypotheses, where the signal-plus-background likelihood (given the prior) is evaluated at the $\Theta$ value that maximizes the likelihood. The distribution of negative log-likelihood-ratio (LLR) values is then used to compute the $p$-value by calculating the fraction of PEs that have a more signal-like LLR value than the observed LLR value in data. The $p$-value is the probability of observing an excess, at least as signal-like as the one observed in data, given that only background exists.


\section{Results}

Good agreement is observed between the data and expected background yields. The most significant deviation from the expected background is seen in the dimuon channel for the CI search, with a $p$-value of 8\% in the LL model with destructive interference given the 1/$\Lambda^2$ prior.  In the ADD search, the most significant excess is also observed in the dimuon channel, with a $p$-value of 6\% in the GRW formalism for the 1/$M_{\rm S}^{4}$ prior.  In neither case is the deviation significant. The expected and observed 95\% C.L. lower exclusion limits are set on the parameter of interest in each search, with the resulting limits for the CI and ADD search presented in Tables~\ref{ExpObs_Limits_CI_LLLRRR} and \ref{tab_ADD_results_formalisms} respectively, including conversions to other formalisms. These results are also displayed graphically in Figure~\ref{CISummary} for the CI search given the 1/$\Lambda^2$ prior and Figure~\ref{ADDSummary} for the ADD search given the 1/$M_{\rm S}^8$ prior.
In the case of the ADD interpretation, the limits obtained with a prior uniform and positive
in signal cross-section are found to be consistent with those obtained with the
uniform positive 1/$M_{\rm S}^8$ prior.

For the ADD search results, the similar expected and observed exclusion limits within the separate channels are due to the small number of expected SM background events, which arise from the high mass threshold chosen for that search.  This leads a large fraction of the PEs to return a result of zero expected events, and the median value of the ensemble (taken as the expected limit) to therefore also be zero expected events. For the combined dilepton channel, the total number of expected SM background events is large enough that a wider range of limits is obtained in the ensemble of PEs and the slight data deficit translates into stronger observed limits than expected.

\begin{table*}[htbp]
\newcommand\T{\rule{0pt}{2.6ex}}
\newcommand\B{\rule[-1.2ex]{0pt}{0pt}}
\caption{Expected and observed 95\% C.L. lower exclusion limits on $\Lambda$ for the LL, LR, and RR contact interaction search using a uniform positive prior in 1/$\Lambda^2$ or 1/$\Lambda^4$. The dielectron, dimuon, and combined dilepton channel limits are shown for both the constructive and destructive interference cases.}
\label{ExpObs_Limits_CI_LLLRRR}
\vspace{0.2cm}
\centering
\begin{tabular}{c|c|c|c|c|c|c|c}
\hline
\hline
\multicolumn{8}{c}{Expected and observed lower limits on $\Lambda$ [\TeV{}]} \T\B \\
\hline
\multirow{2}{*}{Channel} & \multirow{2}{*}{Prior} & \multicolumn{2}{c|}{Left-Left} \T & \multicolumn{2}{c|}{Left-Right} & \multicolumn{2}{c}{Right-Right} \\
\cline{3-8}
& & Const. \T & Destr. & Const. & Destr. & Const. & Destr. \\
\hline
\hline
Exp: $ee$ \T & \multirow{2}{*}{1/$\Lambda^2$} & 19.1 & 14.0 & 22.0 & 17.4 & 19.0 & 14.2 \\
Obs: $ee$    & & 20.7 & 16.4 & 25.2 & 19.2 & 20.2 & 16.6 \\
\hline
Exp: $ee$ \T & \multirow{2}{*}{1/$\Lambda^4$} & 17.4 & 13.0 & 20.1 & 16.3 & 17.2 & 13.1 \\
Obs: $ee$    & & 18.6 & 14.7 & 22.2 & 17.7 & 18.3 & 14.9 \\
\hline
\hline
Exp: $\mu\mu$ \T & \multirow{2}{*}{1/$\Lambda^2$} & 18.0 & 12.7 & 21.6 & 16.3 & 17.7 & 13.0 \\
Obs: $\mu\mu$    & & 16.7 & 12.5 & 20.5 & 14.9 & 16.5 & 12.7 \\
\hline
Exp: $\mu\mu$ \T & \multirow{2}{*}{1/$\Lambda^4$} & 16.2 & 12.0 & 19.8 & 15.3 & 16.2 & 12.1 \\
Obs: $\mu\mu$    & & 15.6 & 11.8 & 19.0 & 14.3 & 15.4 & 11.9 \\
\hline
\hline
Exp: $\ell\ell$ \T & \multirow{2}{*}{1/$\Lambda^2$} & 21.4 & 14.7 & 24.8 & 18.5 & 21.0 & 15.0 \\
Obs: $\ell\ell$    & & 21.6 & 17.2 & 26.3 & 19.0 & 21.1 & 17.5 \\
\hline
Exp: $\ell\ell$ \T & \multirow{2}{*}{1/$\Lambda^4$} & 19.1 & 13.8 & 23.1 & 17.6 & 19.1 & 14.2 \\
Obs: $\ell\ell$    & & 19.6 & 15.4 & 23.8 & 17.8 & 19.3 & 15.6 \\
\hline
\hline
\end{tabular}
\end{table*}

\begin{table*}[htbp]
\newcommand\T{\rule{0pt}{2.6ex}}
\newcommand\B{\rule[-1.2ex]{0pt}{0pt}}
  \begin{center}
    \caption{Expected and observed 95\% C.L. lower exclusion limits on $M_{\rm S}$, using a uniform positive prior in $1/M_{\rm S}^4$ or $1/M_{\rm S}^8$. The dielectron, dimuon, and combined dilepton channel limits are shown for ADD signal in the GRW, Hewett and HLZ formalisms.}
    \label{tab_ADD_results_formalisms}
    \vspace{0.2cm}
\begin{tabular}{c|c|c|c|cccccc}
\hline
\hline
\multicolumn{10}{c}{Expected and observed lower limits on $M_{\rm S}$ [\TeV{}]} \T\B \\
\hline
\multirow{2}{*}{Channel} & \multirow{2}{*}{Prior} & \multirow{2}{*}{GRW} & \multirow{2}{*}{Hewett} & \multicolumn{6}{c}{HLZ} \T \\
\cline{5-10}
        &       &     &        & $n=2$ \T & $n=3$ & $n=4$ & $n=5$ & $n=6$ & $n=7$ \\
\hline
\hline
Exp: $ee$ \T   & \multirow{2}{*}{$1/M_{\rm S}^{4}$} & 4.0 & 3.5 & 3.6 & 4.7 & 4.0 & 3.6 & 3.3 & 3.1 \\
Obs: $ee$      &  & 4.0 & 3.5 & 3.6 & 4.7 & 4.0 & 3.6 & 3.3 & 3.1 \\
\hline
Exp: $ee$ \T   & \multirow{2}{*}{$1/M_{\rm S}^{8}$} & 3.7 & 3.3 & 3.1 & 4.4 & 3.7 & 3.4 & 3.1 & 3.0 \\
Obs: $ee$      &  & 3.7 & 3.3 & 3.1 & 4.4 & 3.7 & 3.4 & 3.1 & 3.0 \\
\hline
\hline
Exp: $\mu\mu$ \T   & \multirow{2}{*}{$1/M_{\rm S}^{4}$} & 3.7 & 3.3 & 3.4 & 4.4 & 3.7 & 3.4 & 3.1 & 3.0 \\
Obs: $\mu\mu$      &  & 3.7 & 3.3 & 3.4 & 4.4 & 3.7 & 3.4 & 3.1 & 3.0 \\
\hline
Exp: $\mu\mu$ \T   & \multirow{2}{*}{$1/M_{\rm S}^{8}$} & 3.5 & 3.1 & 3.1 & 4.2 & 3.5 & 3.2 & 3.0 & 2.8 \\
Obs: $\mu\mu$      &  & 3.5 & 3.1 & 3.1 & 4.2 & 3.5 & 3.2 & 3.0 & 2.8 \\
\hline
\hline
Exp: $\ell\ell$ \T   & \multirow{2}{*}{$1/M_{\rm S}^{4}$} & 4.0 & 3.6 & 3.9 & 4.8 & 4.0 & 3.6 & 3.4 & 3.2 \\
Obs: $\ell\ell$      &  & 4.2 & 3.8 & 4.2 & 5.0 & 4.2 & 3.8 & 3.6 & 3.4 \\
\hline
Exp: $\ell\ell$ \T   & \multirow{2}{*}{$1/M_{\rm S}^{8}$} & 3.8 & 3.4 & 3.5 & 4.6 & 3.8 & 3.5 & 3.2 & 3.1 \\
Obs: $\ell\ell$      &  & 4.0 & 3.6 & 3.7 & 4.7 & 4.0 & 3.6 & 3.4 & 3.2 \\
\hline
\hline
\end{tabular}
  \end{center}
\end{table*}

\begin{figure*}[htb!]
 \begin{center}
  \includegraphics[width=0.85\textwidth]{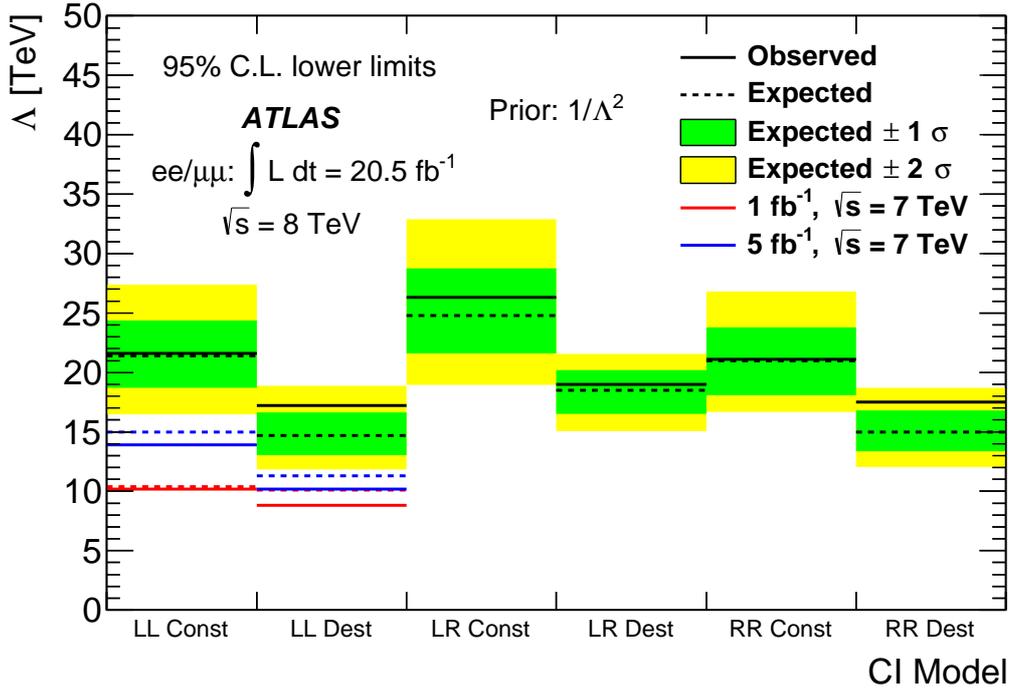}
 \end{center}
 \caption{Summary of 95\% C.L. lower exclusion limits on $\Lambda$ for the combined dilepton contact interaction search, using a uniform positive prior in 1/$\Lambda^2$. Previous ATLAS search results~\cite{Aad:2012bsa, Aad:2011tq} are also presented for comparison. Exclusion limits were previously only set on the LL model.}
 \label{CISummary}
\end{figure*}

\begin{figure*}[htb!]
 \begin{center}
  \includegraphics[width=0.85\textwidth]{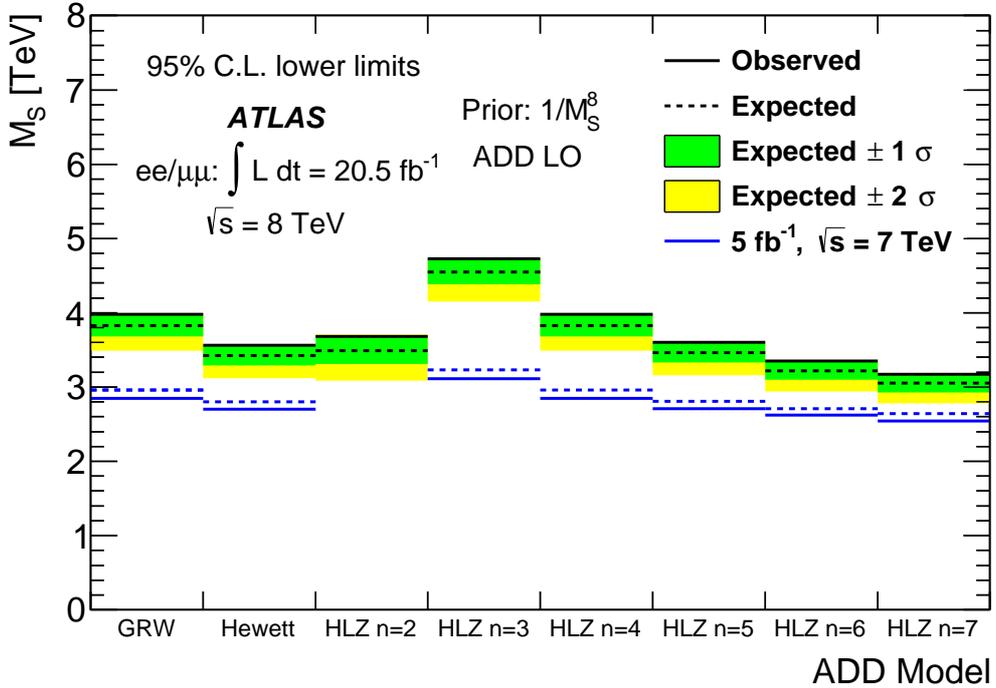}
 \end{center}
 \caption{Summary of 95\% C.L. lower exclusion limits on $M_{\rm S}$ for the combined dilepton ADD large extra dimensions search, using a uniform positive prior in 1/$M_{\rm S}^8$. Previous ATLAS search results~\cite{Aad:2012bsa} are also presented for comparison. Exclusion limits were not previously set on the HLZ $n=2$ ADD model.}
 \label{ADDSummary}
\end{figure*}


\section{Conclusions}

A search for non-resonant new phenomena in the dilepton channel has been
carried out using the 2012 LHC proton--proton collision dataset of 20~\ifb{} 
at $\sqrt{s} = 8$~\TeV{}. This study builds upon previous ATLAS searches, using both dilepton invariant mass and the lepton $\cos\theta^*$ distribution (and by proxy $A_{\rm FB}$) as search variables.  No significant deviations from the Standard Model predictions are observed and lower limits are placed on the scale of contact interactions and large extra dimensions. The most restrictive 95\% C.L. limits are obtained by combining the dielectron and dimuon channels, yielding $\Lambda > 26.3$~\TeV{} for the left-right contact interaction model with constructive interference and a prior flat in 1/$\Lambda^2$, and $M_{\rm S} > 5.0$~\TeV{} for the HLZ $n=3$ ADD model with a prior flat in 1/$M_{\rm S}^4$.


\section*{Acknowledgements}

We thank CERN for the very successful operation of the LHC, as well as the
support staff from our institutions without whom ATLAS could not be
operated efficiently.
We thank Tobias Pook for help with the ADD model.

We acknowledge the support of ANPCyT, Argentina; YerPhI, Armenia; ARC,
Australia; BMWF and FWF, Austria; ANAS, Azerbaijan; SSTC, Belarus; CNPq and FAPESP,
Brazil; NSERC, NRC and CFI, Canada; CERN; CONICYT, Chile; CAS, MOST and NSFC,
China; COLCIENCIAS, Colombia; MSMT CR, MPO CR and VSC CR, Czech Republic;
DNRF, DNSRC and Lundbeck Foundation, Denmark; EPLANET, ERC and NSRF, European Union;
IN2P3-CNRS, CEA-DSM/IRFU, France; GNSF, Georgia; BMBF, DFG, HGF, MPG and AvH
Foundation, Germany; GSRT and NSRF, Greece; ISF, MINERVA, GIF, I-CORE and Benoziyo Center,
Israel; INFN, Italy; MEXT and JSPS, Japan; CNRST, Morocco; FOM and NWO,
Netherlands; BRF and RCN, Norway; MNiSW and NCN, Poland; GRICES and FCT, Portugal; MNE/IFA, Romania; MES of Russia and ROSATOM, Russian Federation; JINR; MSTD,
Serbia; MSSR, Slovakia; ARRS and MIZ\v{S}, Slovenia; DST/NRF, South Africa;
MINECO, Spain; SRC and Wallenberg Foundation, Sweden; SER, SNSF and Cantons of
Bern and Geneva, Switzerland; NSC, Taiwan; TAEK, Turkey; STFC, the Royal
Society and Leverhulme Trust, United Kingdom; DOE and NSF, United States of
America.

The crucial computing support from all WLCG partners is acknowledged
gratefully, in particular from CERN and the ATLAS Tier-1 facilities at
TRIUMF (Canada), NDGF (Denmark, Norway, Sweden), CC-IN2P3 (France),
KIT/GridKA (Germany), INFN-CNAF (Italy), NL-T1 (Netherlands), PIC (Spain),
ASGC (Taiwan), RAL (UK) and BNL (USA) and in the Tier-2 facilities
worldwide.


\bibliographystyle{modspphys}
\bibliography{nonres_paper}

\onecolumn
\clearpage
\input{atlas_authlist}


\end{document}
\endinput

%% file: atlas_authlist.tex
\begin{flushleft}
{\Large The ATLAS Collaboration}

\bigskip

G.~Aad$^{\rm 84}$,
B.~Abbott$^{\rm 112}$,
J.~Abdallah$^{\rm 152}$,
S.~Abdel~Khalek$^{\rm 116}$,
O.~Abdinov$^{\rm 11}$,
R.~Aben$^{\rm 106}$,
B.~Abi$^{\rm 113}$,
M.~Abolins$^{\rm 89}$,
O.S.~AbouZeid$^{\rm 159}$,
H.~Abramowicz$^{\rm 154}$
% ...
\bigskip
\\
% ...
$^{11}$ Institute of Physics, Azerbaijan Academy of Sciences, Baku, Azerbaijan\\
% ...
$^{84}$ CPPM, Aix-Marseille Universit{\'e} and CNRS/IN2P3, Marseille, France\\
% ...
$^{89}$ Department of Physics and Astronomy, Michigan State University, East Lansing MI, United States of America\\
% ...
$^{106}$ Nikhef National Institute for Subatomic Physics and University of Amsterdam, Amsterdam, Netherlands\\
% ...
$^{112}$ Homer L. Dodge Department of Physics and Astronomy, University of Oklahoma, Norman OK, United States of America\\
$^{113}$ Department of Physics, Oklahoma State University, Stillwater OK, United States of America\\
% ...
$^{116}$ LAL, Universit{\'e} Paris-Sud and CNRS/IN2P3, Orsay, France\\
% ...
$^{152}$ Institute of Physics, Academia Sinica, Taipei, Taiwan\\
% ...
$^{154}$ Raymond and Beverly Sackler School of Physics and Astronomy, Tel Aviv University, Tel Aviv, Israel\\
% ...
$^{159}$ Department of Physics, University of Toronto, Toronto ON, Canada\\
% ...
\end{flushleft}